\documentstyle[epsf]{mn}

\newif\ifAMStwofonts

\def\dist{(m-M)_0}
\def\logz{\lbrack\hbox{M/H}\rbrack}
\def\feh{\lbrack\hbox{Fe/H}\rbrack}
\newcommand{\mean}[1]{\langle #1 \rangle}

\def\eps@scaling{.95}
\def\epsscale#1{\gdef\eps@scaling{#1}}
\def\plotone#1{\centering \leavevmode
    \epsfxsize=\eps@scaling\columnwidth \epsfbox{#1}}

\ifoldfss
  \ifCUPmtlplainloaded \else
    \NewTextAlphabet{textbfit} {cmbxti10} {}
    \NewTextAlphabet{textbfss} {cmssbx10} {}
    \NewMathAlphabet{mathbfit} {cmbxti10} {} 
    \NewMathAlphabet{mathbfss} {cmssbx10} {} 
  \fi
  \ifAMStwofonts
    \ifCUPmtlplainloaded \else
      \NewSymbolFont{upmath} {eurm10}
      \NewSymbolFont{AMSa} {msam10}
      \NewMathSymbol{\upi}     {0}{upmath}{19}
      \NewMathSymbol{\umu}     {0}{upmath}{16}
      \NewMathSymbol{\upartial}{0}{upmath}{40}
      \NewMathSymbol{\leqslant}{3}{AMSa}{36}
      \NewMathSymbol{\geqslant}{3}{AMSa}{3E}

      \let\geq=\geqslant 
    \fi
  \fi
\fi 

\ifnfssone
  \newmathalphabet{\mathit}
  \addtoversion{normal}{\mathit}{cmr}{m}{it}
  \addtoversion{bold}{\mathit}{cmr}{bx}{it}
  \newmathalphabet{\mathbfit} 
  \addtoversion{normal}{\mathbfit}{cmr}{bx}{it}
  \addtoversion{bold}{\mathbfit}{cmr}{bx}{it}
  \newmathalphabet{\mathbfss} 
  \addtoversion{normal}{\mathbfss}{cmss}{bx}{n}
  \addtoversion{bold}{\mathbfss}{cmss}{bx}{n}
  \ifAMStwofonts
    \ifCUPmtlplainloaded \else
      \UseAMStwoboldmath
      \makeatletter
      \new@mathgroup\upmath@group
      \define@mathgroup\mv@normal\upmath@group{eur}{m}{n}
      \define@mathgroup\mv@bold\upmath@group{eur}{b}{n}
      \edef\UPM{\hexnumber\upmath@group}
      \new@mathgroup\amsa@group
      \define@mathgroup\mv@normal\amsa@group{msa}{m}{n}
      \define@mathgroup\mv@bold\amsa@group{msa}{m}{n}
      \edef\AMSa{\hexnumber\amsa@group}
      \makeatother
      \mathchardef\upi="0\UPM19
      \mathchardef\umu="0\UPM16
      \mathchardef\upartial="0\UPM40
      \mathchardef\leqslant="3\AMSa36
      \mathchardef\geqslant="3\AMSa3E

      \let\geq=\geqslant 
    \fi
  \fi
\fi 

\ifnfsstwo
  \DeclareMathAlphabet{\mathbfit}{OT1}{cmr}{bx}{it}
  \SetMathAlphabet\mathbfit{bold}{OT1}{cmr}{bx}{it}
  \DeclareMathAlphabet{\mathbfss}{OT1}{cmss}{bx}{n}
  \SetMathAlphabet\mathbfss{bold}{OT1}{cmss}{bx}{n}
  \ifAMStwofonts
    \ifCUPmtlplainloaded \else
      \DeclareSymbolFont{UPM}{U}{eur}{m}{n}
      \SetSymbolFont{UPM}{bold}{U}{eur}{b}{n}
      \DeclareSymbolFont{AMSa}{U}{msa}{m}{n}
      \DeclareMathSymbol{\upi}{0}{UPM}{"19}
      \DeclareMathSymbol{\umu}{0}{UPM}{"16}
      \DeclareMathSymbol{\upartial}{0}{UPM}{"40}
      \DeclareMathSymbol{\leqslant}{3}{AMSa}{"36}
      \DeclareMathSymbol{\geqslant}{3}{AMSa}{"3E}

      \let\geq=\geqslant 
    \fi
  \fi
\fi 

\ifCUPmtlplainloaded \else
  \ifAMStwofonts \else 
    \def\upi{\pi}
    \def\umu{\mu}
    \def\upartial{\partial}
  \fi
\fi

\title[Star Formation History Measurement]{Numerical Methods of Star Formation History Measurement and Applications to Seven Dwarf Spheroidals}
\author[A. E. Dolphin]
 {A. E. Dolphin,$^1$ \\
$^1$Kitt Peak National Observatory, National Optical Astronomy Observatories, PO Box 26732, Tucson, AZ, 85726, USA \\
electronic mail: dolphin@noao.edu}
\date{Accepted . Received }
\pagerange{\pageref{firstpage}--\pageref{lastpage}}
\pubyear{2001}

\begin{document}
\label{firstpage}

\maketitle

\begin{abstract}
A comprehensive study of the measurement of star formation histories from colour-magnitude diagrams (CMDs) is presented, with an emphasis on a variety of subtle issues involved in the generation of model CMDs and maximum likelihood solution.  Among these are the need for a complete sampling of the synthetic CMD, the use of of proper statistics for dealing with Poisson-distributed data (and a demonstration of why $\chi^2$ must not be used), measuring full uncertainties in all reported parameters, quantifying the goodness-of-fit, and questions of binning the CMD and incorporating outside information.  Several example star formation history measurements are given.  Two examples involve synthetic data, in which the input and recovered parameters can be compared to locate possible flaws in the methodology (none were apparent) and measure the accuracy with which ages, metallicities, and star formation rates can be recovered.  Solutions of the histories of seven Galactic dwarf spheroidal companions (Carina, Draco, Leo I, Leo II, Sagittarius, Sculptor, and Ursa Minor) illustrate the ability to measure star formation histories given a variety conditions -- numbers of stars, complexity of star formation history, and amount of foreground contamination.  Significant measurements of ancient $> 8$ Gyr star formation are made in all seven galaxies.  Sculptor, Draco, and Ursa Minor appear entirely ancient, while the other systems show varying amounts of younger stars.
\end{abstract}


\section{INTRODUCTION}

The evolution of galaxies can be studied two ways -- one can either look at high redshift to observe the past directly, or one can look at the fossil remains of past events in nearby galaxies.  These approaches are complementary, as the first is more direct (the ancient light is being observed now) but allows only a statistical comparison of the events seen happening at different ages (we cannot be entirely sure which systems at high redshift are analogous to which systems in the nearby universe).  In contrast the measurement of the star formation history of a nearby galaxy (one whose stellar content is resolved) allows one to trace the history of a single system, but it is difficult to determine that history in an unambiguous way.

The measurement of star formation histories via comparisons of observed and synthetic colour-magnitude diagrams (CMDs) is an active field that is evolving rapidly.  The first papers on the topic arrived in the literature only slightly more than a decade ago, with Tosi et al. (1989) and Bertelli et al. (1992) two early attempts to derive the star formation histories of composite populations (stars of a range of ages and metallicities).  As opposed to isochrone fitting in single-population objects, such as globular clusters, the measurement of a star formation history of a composite system is a daunting task -- SFR(t,Z), distance, extinction/reddening, initial mass function (IMF), and binary distribution are all unknowns at some level; while the comparison of CMDs was done subjectively.  In order to cope with the vast combinations of parameters possible given limitations in computer speed and the number of subjective comparisons that could be made, these early studies limited the parameter space, generally measuring only a small set of SFR(t) functions and assuming fixed values for all other parameters.

The work of Gallart et al. (1996a), studying the old stellar content of NGC 6822, was the first attempt to quantify the subjective CMD comparisons.  In that work, the authors constructed a large number of parameters, each of which measured the position, size, and/or number of stars of a certain feature of the CMD.  This allowed the first quantitative judgment of a star formation history, although both the procedures for generating synthetic CMDs and comparing CMDs were still extremely slow, forcing a solution of only SFR(t) and Z(t).

The shift to a fully quantitative analysis was proposed independently by Dolphin (1997) and Aparicio, Gallart, \& Bertelli (1997), who proposed binning the CMD into sections and performing a $\chi^2$ minimization on the number of stars in each section to determine the star formation history.  Dolphin (1997) demonstrated the sensitivity of such a method to metallicity, distance, and extinction, and as well as its ability to correctly reconstruct the star formation history of a synthetic population; Aparicio et al. (1997) applied a remarkably similar algorithm to a study of LGS 3.  The advantage in such a technique lay in its ability to use all parts of the CMD in measuring the star formation history -- thus allowing it to be used on photometry of any quality and depth -- as well as the obvious advantage of having a single-parameter fit that can be used in a numerical minimization.

The number of groups working on this topic continues to increase; Tolstoy \& Saha (1996); Holtzman et al. (1999); Olsen (1999); Hernandez, Gilmore, \& Valls-Gabaud (2000); and Harris \& Zaritsky (2001) is only a partial list of other groups that are using CMDs to measure past star formation histories.  These techniques have been applied to many of the Local Group galaxies, as well as a few galaxies just outside the Local Group.  Despite the large number of papers on this topic, the literature lacks thorough methodology papers describing modern techniques, largely because of the incremental improvements in methods that have been implemented by each group.  An example of this is the series of papers by Dolphin -- Dolphin (1997) presenting the initial method; Dolphin (2000a), Dolphin et al. (2001a), Miller et al. (2001), and Dolphin (2001) each containing minor improvements to the technique -- which forces the reader to follow a paper trail to determine what any one group is currently doing.  Another significant void in the literature is a realistic estimation of how well one can measure star formation histories under a variety of conditions -- number of stars in the CMD, amount of foreground contamination, and complexity of star formation.

The present work attempts to fill these needs in the literature, in addition to addressing commonly-made mistakes.  This paper is divided into two main sections -- a detailed description of how to measure star formation histories from CMDs and application to artificial and real data.

\section{ANALYSIS PROCEDURE}

In any measurement of the star formation history, one fundamental question must be addressed: what set of star formation histories could have created the observations?  In order to answer this question, the following steps must be taken:
\begin{itemize}
\item Generate synthetic CMD based on theoretical isochrones
\item Account for incompleteness and observational errors
\item Measure the best star formation history and its quality
\item Measure the allowable range of other star formation histories
\end{itemize}
Each of these steps will be addressed in the sections below.

\subsection{Synthetic CMD Generation \label{sec_synthcmd}}

The generation of synthetic CMDs, as described in this paper, is a two-step process -- generation of ``clean'' CMDs and the introduction of incompleteness and observational errors.  (The reason for the split is merely a computational one.  Generation of the clean CMDs is the most time-consuming part of the entire measurement, but only needs to be done once; application of observational errors will be different at different assumed distance and extinction values.)  The end result of this process is intended to be a model CMD -- the probability distribution from which the observed data are drawn.  The question of binning the CMD vs. storing individual stars will be addressed in section \ref{sec_fitparam}.  For the time being, we will simply assume that the CMD is to be binned.

As was pointed out by Dolphin (1997), a CMD of a composite population is simply the sum of the CMDs of its constituent parts.  Thus, for any given distance, extinction, IMF, and binary distribution, the CMD corresponding to any SFR(t,Z) can be computed as the sum of its parts.  (If one wishes to solve also for distance or any of the other ``fixed'' parameters, separate solutions must be made at each combination of fixed parameters)  This makes it unnecessary to spend the vast computational resources used in early studies, as the ``partial CMDs'' -- model CMDs containing small ranges in age and metallicity -- need to be computed only once.  If one computes each partial CMD with the same star formation rate, such as $1 M_{\odot} yr^{-1}$, the model CMD for an arbitrary star formation history is given by
\begin{equation}
\label{eq_mi_simple}
m_i = \sum_j r_j c_{i,j},
\end{equation}
where $m_i$ is the full model CMD in bin $i$, $r_j$ is the star formation rate for partial CMD $j$ in $M_{\odot} yr^{-1}$, and $c_{i,j}$ is bin $i$ of partial CMD $j$.  This relation makes the computational problem much easier, but determining $c_{i,j}$ is nevertheless a non-trivial procedure.  The usual procedure for this process is to randomly populate each partial CMD with a large number of stars randomly drawn from the age and metallicity range, and an adopted IMF.  Although an attractive algorithm for its simplicity, it is impossible in practice to adequately sample the model CMD this way as the density of points on the CMD varies by many orders of magnitude between the lower main sequence and the Hertzsprung gap.  Additionally, such a ``random drawing'' routine inevitably adds random errors to the CMD, thus making the CMD comparison one of data-data rather than data-model.  Since we can determine the underlying model via the process described here, there is no need to apply a statistically-weaker data-data comparison that assumes no knowledge of the model.

Thus we seek to find an algorithm that will generate a true model CMD.  In order to accomplish this, one must completely sample all possible combinations of mass (including secondary mass in unresolved binary systems), metallicity and age comprising the partial CMD.  The process will thus first involve calculating a sufficiently large number of isochrones, so that the space between adjacent isochrones is much smaller than the CMD bin size.  For example, the Girardi et al. (2000) isochrones with $(Z,\log t)$ of $(0.001,8.50)$ and $(0.001,8.55)$ have a maximum separation of $\Delta V = 0.45$ magnitudes and $\Delta (V-I) = 0.10$.  Even if using a coarse binning size of $0.1$ magnitudes in $V$, this would require small steps of approximately $\Delta \log t = 0.005$.  The isochrones at $(0.001,8.50)$ and $(0.004,8.50)$ have maximum separations of $\Delta V = 1.04$ and $\Delta (V-I) = 0.71$, thus requiring small steps in metallicity ($\sim 0.02$ dex) to adequately fill in the CMD.  These values are only samples; the actual step sizes used in model generation depend on the spacing between isochrones and the CMD binning size.  It should be noted that one cannot hope to actually measure ages and metallicities at this level of precision; however the presence of those isochrones is necessary in order to provide a complete model CMD.  It is clear that an interpolation scheme is mandatory in this process; the details of this are beyond the scope of this paper.

Once the set of needed isochrones has been established, each isochrone is then considered in turn and divided into an appropriate number of points.  Again, the stepsize is a function of the CMD binning size.  Each point is weighted by the IMF, the mass difference between it and the adjacent points, and the stepsizes in age and metallicity.  Binaries are also added at this point; again it is necessary to sample the range of secondary masses to at least the accuracy of the binned CMD.

The result of this process can be thought of as a ``blurred isochrone'', centred roughly on the central age and metallicity used for the partial CMD.  What is critical is that all possible masses, metallicities, ages, and binary combinations within this range are accounted for in the binned CMD; this goal can only be achieved by the procedure outlined above.

\subsection{Simulating Observational Conditions \label{sec_obserror}}

Of course, an observed CMD is never purely a pure isochrone; photometric errors, blending, incompleteness, bad/false detections, and foreground contamination all complicate matters.  In generating an accurate model CMD, all of these factors must be taken into account.  As was pointed out by Gallart et al. (1996b), the problems of photometric errors, blending, and incompleteness can be addressed in one step through the use of a library of artificial star tests.  Indeed, this is the \textit{only} accurate way of addressing these problems, as incompleteness is a function of the observed magnitude of a star rather than its true magnitude, blending errors depend on the density of stars and the relative distributions within the CMD, and even simple photometric errors are biased and non-Gaussian.

Thus the necessary procedure for simulating the first three observational effects is to generate a very large library of artificial stars, which includes the necessary range of input magnitudes and colours with a sufficiently large number of stars input at each location on the CMD so that the distribution of recovered photometry is adequately sampled.  For each point on the partial CMD created in section \ref{sec_synthcmd}, it is necessary to multiply its weight by the completeness fraction and distribute that weight according to the distribution of recovered artificial stars.

It is also possible to correct for foreground contamination in a consistent manner.  Again it is necessary to realize that foreground stars in the observed data are randomly drawn from an intrinsic distribution in the same manner as the object stars.  Thus they can be modeled in the same manner as the object stars -- by constructing a model foreground CMD.  This is usually not done with isochrones; the common procedure is instead to observe a second field nearby (in terms of Galactic coordinates) the object field but well beyond the limits of the object being studied.  A small amount of smoothing of the foreground CMD is generally necessary, and the resulting CMD can be added to the partial CMDs to create a better representation of the model from which the observed data are drawn.  This procedure is clearly superior to the more commonly-used ``statistical subtraction'', as the subtraction process inevitably leaves residuals (oversubtraction and undersubtraction) and thus a CMD that is \textit{not} representative of the underlying star formation history.  It will be demonstrated in Section \ref{sec_sag} that, when treated properly, foreground contamination can be dealt with easily.

A final problem is the presence of bad points -- short-period variables, stars hit by cosmic rays in one image, bad pixels, etc. -- that cannot be modeled either by the partial CMDs or by a foreground CMD.  Again, it is necessary to attempt to create the underlying model distribution from which these ``objects'' are drawn.  This model consists largely of two distributions.  Purely artificial objects, such as cosmic rays and bad pixels, are likely to be spread anywhere on the observed CMD, and should be fit with a flat distribution.  Short-period variables and stars plus cosmic rays will likely fall near the observed distribution of points, and should be modeled by smoothing either the observed or model CMD (usually the observed CMD).  As with the foreground CMD, the ``bad point'' CMD is to be added to the partial CMDs when determining the model CMD for a star formation history.  Thus equation \ref{eq_mi_simple} becomes
\begin{equation}
\label{eq_mi}
m_i = \sum_j r_j c_{i,j} + f_i + bp_i,
\end{equation}
where $f_i$ is bin $i$ of the foreground CMD and $bp_i$ is bin $i$ of the combined bad point CMD.

\subsection{Comparison of Model CMD with Data \label{sec_fitparam}}

When faced with the task of fitting data to a model, most scientists tend to first think of using $\chi^2$.  As $\chi^2$ is simply related to the differences between data and model and the predicted $1 \sigma$ errors, there is a certain intuitive attractiveness to this approach.  Less appreciated, though, is the fact that minimizing $\chi^2$ is actually a maximum-likelihood calculation for the case of data with Gaussian errors and known uncertainties at each point.  This can be demonstrated trivially as follows.  $P_i$ denotes the probability that the observation $n$ is drawn from model $m$, $m_i$ is the model value of bin $i$, $n_i$ is observed value of bin $i$, and $\sigma_i$ is the uncertainty of bin $i$.
\begin{equation}
P_i = \sqrt{\frac{1}{2 \pi {\sigma_i^2}}} e^{- 0.5 (n_i - m_i)^2/{\sigma_i^2}}
\end{equation}
We can define a ``Gaussian likelihood ratio'' as the probability that observed data point $n_i$ was drawn from a model equal to $m_i$ divided by the probability that it was drawn from a model equal to $n_i$.  (This is equivalent to the term ``relative probability'' used by Tolstoy \& Saha 1996).
\begin{equation}
\label{eq_glri}
GLR_i = \sqrt{\frac{\sigma_{ni}^2}{{\sigma_{mi}^2}}} e^{- 0.5 (n_i - m_i)^2/{\sigma_{mi}^2}},
\end{equation}
where $\sigma_{mi}$ equals the expected uncertainty with model $m_i$ and $\sigma_{ni}$ is that for model $n_i$.  Multiplying the individual Gaussian likelihood ratios and taking the logarithm, we obtain
\begin{equation}
-2 \ln GLR = \sum_i \ln \frac{\sigma_{mi}^2}{\sigma_{ni}^2} + \sum_i \frac{(n_i - m_i)^2}{\sigma_{mi}^2},
\end{equation}
or simply
\begin{equation}
-2 \ln GLR = \chi^2 + \sum_i \ln \frac{\sigma_{mi}^2}{\sigma_{ni}^2}.
\end{equation}
Thus, if the observational error distribution is a smooth Gaussian, and if the $\sigma_i$ values do not change during the fit (in which case $\ln ( \sigma_{mi}^2 / \sigma_{mi}^2 ) = 0$), minimizing $\chi^2$ is the equivalent of finding the model most likely to have produced the observations.  However, neither of these assumptions is true in CMD analysis -- the data follow a Poisson distribution, and $\sigma_{mi}^2 = m_i$ while $\sigma_{ni}^2 = n_i$.

The danger in using $\chi^2$ to minimize Poisson-distributed data is that the determined ``solution'' will not actually be the correct solution.  Examples are given by Mighell (1999); the reader can easily verify this fact by populating an array with, on average, 1 point per bin and minimizing $\chi^2$ to find the mean.  Depending on the formulation of $\chi^2$ used, one's ``fit'' will be incorrect by up to 42\%.  Mighell (1999) proposes an alternative statistic ($\chi^2_{\gamma}$) that will minimize properly, but a better solution can be found by deriving a statistic based on a Poisson, rather than Gaussian, probability function.

Instead of using a $\chi^2$ fit, with its implicit assumptions of the data, one should instead use a maximum likelihood parameter based on the Poisson probability distribution
\begin{equation}
P_i = \frac{m_i^{n_i}}{e^{m_i} n_i!}.
\end{equation}
The ``Poisson likelihood ratio'' is analogous to the Gaussian likelihood ratio ($\chi^2$) in equation \ref{eq_glri}.  Cancelling the $n_i!$ terms in numerator and denominator, we have
\begin{equation}
PLR_i = \frac{m_i^{n_i} e^{n_i}}{n_i^{n_i} e^{m_i}},
\end{equation}
the ratio of the probability of drawing $n_i$ points from model $m_i$ to that of drawing $n_i$ points from model $n_i$.  The cumulative likelihood ratio is given by
\begin{equation}
PLR = \prod_i (\frac{m_i}{n_i})^{n_i} e^{n_i - m_i},
\end{equation}
and the Poisson equivalent of $\chi^2$ is
\begin{equation}
\label{eq_plr}
-2 \ln PLR = 2 \sum_i m_i - n_i + n_i \ln \frac{n_i}{m_i}.
\end{equation}
An examination of this parameter indicates that it shares many of the same features as $\chi^2$, namely that it is zero when $n_i = m_i$ and that the expectation value and variance are 1 and 2, respectively, at large values of $m_i$.  Additionally, minimizing this parameter is truly a maximum likelihood calculation, and applying this parameter to the example given above will result in a correct determination of the mean.  Thus, given the presence of a Poisson-based statistic that can be minimized in the same manner as $\chi^2$, there is no good reason to use $\chi^2$ to fit a CMD, as $\chi^2$ will \textit{always} minimize to the wrong solution.

Before turning our attention to more general aspects of finding the best fit, we need to address a pair of statistics.  First is the Saha $W$ statistic (Saha 1998)
\begin{equation}
W_i = \frac {(m_i+n_i)!}{m_i! n_i!}.
\end{equation}
As noted by Saha (1998), this parameter is proportional to the probability that observed data sets $m_i$ and $n_i$ are drawn from the same model distribution, without any knowledge of what that model is.  It is therefore a data-data comparison and cannot be used for the model-data comparison we wish to perform.  (The fact that one is taking a factorial of a non-integer $m_i$ is the first indication that it is unsuitable for such a task.)  Incidentally, a related statistic can be used for comparison of data with a randomly-drawn synthetic CMD, though it is significantly more complex than the Poisson likelihood ratio.  Rather than determining the likelihood that two data sets are random realizations the same model (the basis of the $W$ statistic), one instead measures the likelihood that the observed data are a random realization of some linear combination of the models from which the synthetic CMDs are drawn.  This probability is given by
\[
P = \prod_i \lbrace \frac{1}{n_i! \prod_j s_{ij}!} \int_{m_{i1} = 0}^{\infty} ... \int_{m_{in} = 0}^{\infty}
\]
\begin{equation}
\lbrack e^{-\sum_j (c_j+1) m_{ij}} ( \sum_j c_j m_{ij})^{n_i} \prod_j (m_{ij}^{s_{ij}} dm_{ij}) \rbrack \rbrace,
\end{equation}
where $n_i$ is the number of observed points in CMD bin $i$, $s_{ij}$ is the number of synthetic points in CMD bin $i$ of partial CMD $j$, $m_{ij}$ is the underlying model in CMD bin $i$ of partial CMD $j$, and $c_j$ is the star formation rate corresponding to partial CMD $j$.  Substituting $x_{ij} = m_{ij} (c_j+1)$ and $y_j = c_j/(c_j+1)$, this simplifies slightly to
\[
P = \prod_i \lbrace \frac{\prod_j (1-y_j)^{s_{ij}}}{n_i! \prod_j s_{ij}!} \int_{x_{i1} = 0}^{\infty} ... \int_{x_{in} = 0}^{\infty}
\]
\begin{equation}
\lbrack ( \sum_j y_j x_{ij})^{n_i} \prod_j (e^{-x_{ij}} x_{ij}^{s_{ij}} dx_{ij}) \rbrack \rbrace.
\end{equation}
Expanding the first sum inside the integral to a polynomial, the integral is reduced to a set of gamma functions, which can be easily solved.

The final statistical treatment to be considered is the Bayesian inference scheme proposed by Tolstoy \& Saha (1996), which is usually seen as a ``bin-free'' statistic.  However, if the CMD binning grid is sufficiently fine, so that $m_i$ adequately describes the density of model points everywhere within the bin (in other words, the binning size is smaller than the CMD features), then their probability of measuring a point in CMD bin $i$ becomes merely the fraction of model points that are in that CMD bin, or
\begin{equation}
P_i = \frac{m_i}{\sum_j m_j}.
\end{equation}
In this equation $P_i$ is the probability of an observed point falling in CMD bin $i$, and $m_i$ is (as before) the number of model points in CMD bin $i$.  This formulation is a slight improvement over their equation 11, as it allows for the more accurate treatment of observational errors described in \ref{sec_obserror} instead of the unbiased Gaussian errors assumed by Tolstoy \& Saha (1996).  The cumulative probability of drawing the entire observed data set $n_i$ is thus given by
\begin{equation}
P = \prod_i (\frac{m_i}{\sum_j m_j})^{n_i},
\end{equation}
which produces
\begin{equation}
\label{eq_bip}
-2 \ln P = \sum_i n_i \ln \frac{m_i}{\sum_j m_j}.
\end{equation}
Aside from factors of the model normalization ($\sum m_i$), equation \ref{eq_bip} will minimize identically to equation \ref{eq_plr} above -- the only difference is that equation \ref{eq_bip} throws away the overall star formation rate information and thus returns only relative star formation rates, while equation \ref{eq_plr} does not.  As there is nothing to be gained by using equation \ref{eq_bip} instead of equation \ref{eq_plr}, we will not discuss the Bayesian inference scheme further.  It should be noted, however, that use of Bayesian inference (as formalized in equation \ref{eq_bip}) is not an ``error'' in the sense that a $\chi^2$ fit is wrong; it simply is of less value than the Poisson likelihood ratio.

\subsection{Determination of the Best Fit}

Something that is frequently considered in minimization solutions is the particular algorithm used to determine the best fit.  Specifically, genetic and annealing algorithms are commonly applied because these are less likely to be trapped in local minima.  However, it is worth considering whether this is actually a valid concern.  Given any arbitrary star formation history $r_j$ that produces a model CMD given by $m_i$, the Poisson likelihood ratio will be given by
\[
fit(r_j) = 2 \sum_i \lbrack \sum_j r_j c_{i,j} + f_i + bp_i \rbrack - n_i
\]
\begin{equation}
+ n_i \ln \frac{n_i}{\sum_j r_j c_{i,j} + f_i + bp_i}.
\end{equation}
If $dv_j$ is a small vector in the direction of the best fit, the Poisson likelihood ratio at $r_j + dv_j$ is given by
\[
fit(r_j + dv_j) = 2 \sum_i \lbrack \sum_j (r_j + dv_j) c_{i,j} + f_i + bp_i \rbrack - n_i
\]
\begin{equation}
+ n_i \ln \frac{n_i}{\sum_j (r_j + dv_j) c_{i,j} + f_i + bp_i}.
\end{equation}
The change in the likelihood ratio by introducing $dv_j$ is
\begin{equation}
2 \sum_i \lbrace \lbrack \sum_j dv_j c_{i,j} \rbrack - n_i \ln \lbrack 1 + \frac {\sum_j dv_j c_{i,j}}{\sum_j r_j c_{i,j} + f_i + bp_i} \rbrack \rbrace.
\end{equation}
Since the vector $dv_j$ is arbitrarily small, we can approximate $\ln (1+x)$ with x, producing
\begin{equation}
\Delta fit = 2 \sum_i \lbrace \lbrack \sum_j dv_j c_{i,j} \rbrack ( 1 - \frac{n_i}{m_i} ) \rbrace.
\end{equation}
Since any movement toward the best fit will lower the model CMD where it is currently overestimated ($\sum_j dv_j c_{i,j} < 0$ where $n_i / m_i < 1$) and raise it where it is underestimated  ($\sum_j dv_j c_{i,j} > 0$ where $n_i / m_i > 1$), $\Delta fit$ will always be negative when moving in the right direction.  Thus there are no local minima in the solution for SFR(t,Z), and any reasonable minimization algorithm can be used.

This is generally accomplished by using a minimization routine, such as frprmn from \textit{Numerical Recipes} (Press et al. 1992) to measure the values of $r_j$ (equation \ref{eq_mi}) that minimize the fit parameter.  Alternately, one can apply the variational calculus technique of Hernandez, Valls-Gabaud, \& Gilmore (1999), although significant additions to their method must be made to adequately deal with observational effects and a function SFR(t,Z) that varies with two parameters.  Since the variational calculus approach assumes SFR(t) to be continuous on a timescale of 0.1 Gyr, and since it is demonstrated in the next section that the high resolution in age and metallicity that is required when using the variational calculus technique actually increases the uncertainties in the measurement, we will adopt the former technique.

Regarding the specific choice of minimization routine, any of the general routines given in \textit{Numerical Recipes} -- amoeba, powell, dfpmin, or frprmn -- will work, given trivial modifications to eliminate negative star formation rates.  Downhill simplex (amoeba) and Powell's method (powell) are the simplest, requiring only the ability to measure the fit parameter given any star formation history.  As noted by Press et al. (1992), Powell's method converges significantly faster than a downhill simplex, and thus is preferred.  The remaining two algorithms are potentially faster, provided that they can be supplied with the fit parameter and derivatives at any arbitrary star formation history and that the measurement of the derivatives takes less time than $N$ computations of the fit parameter ($N$ being the number of partial CMDs).  Using the Poisson likelihood ratio defined in equation \ref{eq_plr} and the model CMD defined in equation \ref{eq_mi}, the derivative is given by
\begin{equation}
\frac{df}{dr_k} = \sum_i \frac{df}{dm_i} \frac{dm_i}{dr_k} = 2 \sum_i ( 1 - \frac{n_i}{m_i} ) c_{i,k}.
\end{equation}
The quantity $1-\frac{n_i}{m_i}$ needs to be calculated only once in each CMD bin (and is the longest part of the calculation), allowing the full set of derivatives to be calculated in little more time than the fit parameter itself.  In terms of the choice between the Davidon-Fletcher-Powell (DFP) algorithm and the Fletcher-Reeves-Polak-Ribiere (FRPR) algorithm, I have found that the DFP algorithm (as implemented by Press et al. 1992) frequently fails to converge, while the FRPR algorithm has no such difficulties; the recommendation is thus for the use of the FRPR algorithm (frprmn in Press et al. 1992).

This algorithm converges very quickly if near the minimum (sufficiently close that the fit parameter becomes quadratic); however it can take some time if far away.  A very fast way to provide a rough initial value is to start with all star formation rates set to zero and incrementally add to the rates whose gradients are the most negative.  This requires scaling the partial CMDs to contain comparable numbers of stars, since otherwise the rates producing the most stars -- rather than those most resembling the observed CMD -- will be filled in this technique.  This scaling concern is also present in frprmn, as it is essentially a sophisticated steepest-descent algorithm.

As both the number of iterations required and the time taken per iteration scale as $N$ in the FRPR algorithm and in the initial seeding algorithm, the total time for convergence scales as $N^2$.  If one is using a very large number of partial CMDs, it may be preferable to sacrifice accuracy for speed.  An algorithm that will give a good (though not excellent) fit to the data while scaling as $N$ is given here.  Beginning with an initial guess (usually of a constant star formation rate), the following two-step iteration procedure is made until sufficient convergence is reached.  The first step is a measurement of the model CMD ($m_i$) using equation \ref{eq_mi}; the second is an updating of the star formation rates $r_j$ using
\begin{equation}
r_j = r_{j 0} \frac{\sum_i c_{i,j} n_i / m_i}{\sum_i c_{i,j}}.
\end{equation}
This is a very crude algorithm, and does not converge to high precision quickly.  However, it reaches moderate levels of convergence fast enough that, for $N > 150$, the speed improvement is significant and outweighs the small amount of accuracy lost.

\subsection{Determination of the Uncertainties \label{sec_unc}}

Once one follows the process above and determines a ``best fit'', the resulting numbers are the star formation rates corresponding to each partial CMD and the overall fit parameter.  Arriving at these values is certainly important in star formation history studies, but two central questions remain unanswered: (1) how far from the best fit is the ``true fit'', and (2) how good is the best fit.  This and the following section address these questions.  The ``true fit'' is defined here as the fit corresponding to the actual star formation history.

The first question is of fundamental importance, because merely quoting the ``best-fitting star formation history'' is useless unless one also provides a measurement of the uncertainties.  In Gaussian data fit using $\chi^2$, for example, we know that the mean $\chi^2$ (not reduced) difference between the underlying model and the best fit equals the number of free parameters in the fit.  (This, of course, is why one uses the number of ``degrees of freedom'' -- the number of measurements minus the number of parameters -- when calculating a reduced $\chi^2$.)

For Poisson-distributed data, however, the expectation value of the Poisson likelihood ratio is not a constant value, but rather varies with the number of model points in each bin.  Assuming that the fit is most driven by the bins contributing the most to the variance of the fit parameter, the mean difference between the fit parameter of the best fit and underlying model will be the sum of the expectation values of the first $N$ fit parameters of the bins with the largest expected variances ($N$ being the number of free parameters in the solution).  This sounds more complex than it is, as the variance and expectation values can be calculated easily for any number of model points.  In practice, this value is generally a little more than $N$, as the expectation values slightly exceed 1.0 where the variances are the highest.  One can therefore approximate the difference as equaling the number of free parameters in the fit; a proper calculation requires calculating the fit parameter expectation value and variance (both are functions of the number of model points) in each CMD bin.

A brief comment regarding the determination of the number of free parameters should be made.  Although, by definition, this should equal the total number of partial CMDs, plus one for any foreground or bad star CMD that was fit, plus one for any ``fixed parameter'' that was fit, the number is generally much smaller.  There are two reasons for this -- the restriction of non-negative star formation rates and the inclusion of partial CMDs that in no way resemble any part of the observed CMD.

The effect of the first can be demonstrated easily.  Fitting 10 Gaussian-distributed points, all with mean values of 0, to the line $y = a + b x$ with no restrictions on $a$ and $b$ returns a mean $\chi^2$ of 8.0 -- 10 points minus 2 free parameters.  Requiring $a \geq 0$ causes the mean $\chi^2$ to equal 8.5, effectively producing 1.5 free parameters since $a < 0$ half the time; the same is true of $b$.  Finally, requiring both $a \geq 0$ and $b \geq 0$ returns a mean $\chi^2$ of 9.32.  The average number of free parameters in this fit (2 if both $a$ and $b$ are positive, 1 if one is positive, and two if neither is positive) equals 0.68; thus the effective number of free parameters in a fit that restricts parameters from being negative is simply equal to the number of positive parameters in the solution.

Any partial CMDs that are completely orthogonal to the observed CMD likewise do not add to the effective number of free parameters.  For example, one can again consider 10 Gaussian-distributed points, and fit to $y = a f_1(x) + b f_2(x)$.  If $f_2(x)$ is zero in the range of $x$ values used in these points, the mean $\chi^2$ is 9.0 even though there are technically two free parameters in the fit.  Since, in a CMD fit, a partial CMD is not zero everywhere, $b$ will be constrained (and forced to zero) by the lack of observed stars where its partial CMD is non-zero, we can again simply ignore any star formation rates that are measured to be zero when counting free parameters.

The presence of nearly-degenerate isochrones, however, does not decrease the effective number of free parameters.  For example, fitting 10 Gaussian-distributed points with $x$ values between 0 and 9 to the curve $y = a + b x + c x^{1.000001}$ returns a mean $\chi^2$ of 7.0, despite the nearly complete degeneracy of the second and third terms ($9^{1.000001} = 9.00002$).  In practice, the parameters $b$ and $c$ are generally determined to be very large and opposite numbers, so the limitation of non-negative star formation rates will cause one to be zero and thus not counted as a free parameter.  In either case, the presence of nearly-degenerate isochrones in the solution requires no additional effort in measuring the effective number of free parameters.

Thus armed with knowledge of the minimized fit parameter and the expected difference between the minimized fit parameter and the fit parameter of the underlying model, one can search all free parameters to determine the range of acceptable values.  Measurement of the uncertainties in determinations of fixed parameters such as distance, extinction, etc. are quite simple, as a separate solution must be made for each combination (because the partial CMDs will be different if any fixed parameters change).  One can determine the best fit at each trial distance, for example, and the range of distances producing fit values within the acceptable range gives the uncertainty in distance.

Measurement of uncertainties of the star formation rate and metallicity, however, requires slightly more work.  The uncertainties come from two sources -- uncertainties in the fixed parameters and acceptable ranges within any one fit -- which must be added in quadrature.  The first source is easy to quantify as one can simply find the extreme values of star formation rates and metallicities in the fits returning acceptable fit parameters.  The second source of uncertainty is more difficult, and requires a trial-and-error testing of the acceptable range.  During such a test, a parameter should be fixed to a variety of values, with the other parameters re-solved to minimize the fit.  By allowing the other parameters to vary, one will find the true uncertainty for that parameter, including the effects of correlated errors (as is usually seen in adjacent isochrones).

An alternate approach to the problem is to use a Monte Carlo test.  The usual method for doing this is to build a large number of simulated observed CMDs, using the best-fit fixed parameters and star formation history and solve for the histories of those CMDs.  Any difference between the average solved values and the input values indicates a bias in the solution, while the scatter indicates the uncertainties.  Such a test is not, in fact, correct in the strictest sense, as one should actually attempt to find the range of \textit{input} parameters that produce the same star formation history as did the observed data.  However, if one's star formation history routine is unbiased (which can be tested by solving simulated data, as is done below in sections \ref{sec_sim1} and \ref{sec_sim2}) the Monte Carlo test should provide a reasonably accurate estimate of the uncertainties.  If the solution is biased (such as if one is using a $\chi^2$ fit), the Monte Carlo test cannot be used reliably.

A final caution should be given against arbitrarily high precision in the recovered star formation history.  Hernandez et al. (2000) measure star formation rates from $0-15$ Gyr with steps of 0.1 Gyr, thus effectively using 150 free parameters in the fit.  For the same objects they studied, I will use 11 age bins in my solutions.  The coarser binning strategy produces smaller uncertainties for two reasons.  First, the acceptable range in the fit parameter increases linearly with the number of free parameters; as the fit parameter essentially goes as the square of errors in the parameters, using 11 free parameters instead of 150 decreases uncertainties by a factor of 3.7.  Second, a bin of 0.1 Gyr can have its star formation rate varied quite severely before generating a bad fit; bins of 1 Gyr or more have much less freedom.  This contributes another factor of $N$ to the uncertainties (not $\sqrt{N}$, since errors in adjacent bins are correlated), giving the coarse (11-bin) fit another factor of $13.6$ improvement in precision, or a total of a factor of 50 improvement in the error bars.  As the solutions presented below are generally $1-2 \sigma$ detections using an 11-bin resolution (thus necessitating yet lower resolution), an accurate measurement of the Hernandez et al. (2000) uncertainties would indicate that each point on their curves has a true uncertainty of at leat 10 times its measured value!

\subsection{Measurement of the Fit Quality \label{sec_quality}}

The final main issue that must be addressed here is that of the goodness-of-fit.  This is an entirely different question than that discussed in the previous section.  The last section discussed the fit parameters of various solutions given one observed data set; this section discusses the fit parameters of various observed data sets drawn from the same solution.  For example, again using the analogy of Gaussian-distributed data, the mean difference of $\chi^2$ between solution producing the best fit and the solution from which the data were drawn equals the number of free parameters in the solution.  However, the variance of $\chi^2$ for many data sets drawn from the same model distribution equals twice the number of degrees of freedom.

The goodness-of-fit can be quantified in at least two ways -- using a percentile or a number of $\sigma$ of error -- either of which is acceptable.  The first test is done by generating a large number of fit parameters for artificial data that were drawn randomly from the best-fitting model.  Since one already knows the number of model points in each CMD bin (determined during the minimization process), this can be done very quickly.  By determining where or if the minimized fit parameter (plus the correction for the number of free parameters, as described in section \ref{sec_unc}) falls within the histogram of random drawings, one can ascertain the goodness of the fit.  Such a technique was used by Hernandez et al. (2000).

The second test can be also done easily, as the expectation values and variances of the fit parameter in each CMD bin were determined as described in section \ref{sec_unc}.  By adding these, a combined expectation value and variance can be determined for the best-fitting model.  To quantify the fit quality, the minimized fit parameter (again corrected for the number of free parameters) can be framed in terms of $\sigma$ away from an ideal fit, using the following definition:
\begin{equation}
Q = \frac{fit\ parameter - expectation\ value}{\sqrt{variance}}.
\end{equation}
If $Q$ is zero, the data represent a typical random drawing from the best model.  If $Q$ is one, the data are $1 \sigma$ worse than a typical random drawing from the best model.  Since most scientists are more familiar with $\chi^2$ values, I also define an ``effective $\chi^2$'':
\begin{equation}
\chi^2_{eff} = 1 + Q \sqrt{2/N},
\end{equation}
where $N$ should equal the number of CMD bins containing either stars or some minimum number of model points.  ($N$ could also be defined as the total number of CMD bins, but a bin with zero model points and zero observed points does not contribute to the Poisson likelihood ratio and thus one could obtain arbitrarily good $\chi^2_{eff}$ values by making a vast CMD.)

\subsection{Binning the CMD}

Until now, the question of the exact technique for CMD binning has not been mentioned, as it was not relevant.  The techniques described earlier -- synthetic CMD generation, the Poisson likelihood ratio, and the methods for determining uncertainties and goodness-of-fit -- are valid for any binning scheme.  The obvious choice of a binning size is sufficiently small that CMD features are not lost, with the bin size comparable to the size of the smallest features to which one wishes to be sensitive; this is also the condition necessary for equivalence between binned and unbinned statistics as noted above.

Occasionally, one will find that the fit needs to be weighted towards large-scale features in order to (for example) place the red giant branch (RGB) in the correct position.  Since the acceptable range of fit parameters (Section \ref{sec_unc}) is a function primarily of the number of free parameters, fit parameters calculated at various binning sizes can be averaged together and treated normally.  (One must account for this when measuring uncertainties and fit quality, of course.)

To consider the effect of two bin sizes, consider the following two examples.  In the first case, there is a $4 \times 4$ block of CMD bins, all with observed values $1 \sigma$ above the model values plus random scatter of $1 \sigma$.  (For simplicity, this discussion will be in terms of Gaussian-distributed data and $\chi^2$; the principles hold true for Poisson-distributed data.)  The $\chi^2$ from this set of bins is $2$ per bin, fit a total of $\chi^2 = 32$ compared with an expectation value of $\chi^2 = 16$ for 16 bins.  In contrast, consider the same block of bins, with half $1 \sigma$ high and half $1 \sigma$ low.  This likewise has a $\chi^2$ of 32.  Now combining the $4 \times 4$ block of bins into a single bin, we re-examine the two cases.  In the first, the $\chi^2$ value is $17$, again 16 higher than the expectation value of $\chi^2 = 1$ for 1 bin.  In the second, however, the $\chi^2$ value is 1, equal to the expectation value.  The conclusion of this exercise is that increasing the bin sizes does not \textit{increase} the sensitivity of the fit parameter to large features; rather it \textit{decreases} the sensitivity of the fit parameter to small features.  This may be necessary in some instances, but it should be done with caution.

The final issue regarding binning schemes is whether to use ``smooth binning'' (dividing the entire CMD into equal-sized rectangular bins) or ``irregular binning'' (selecting specially-shaped bins for different CMD regions).  As mentioned previously, the Poisson likelihood ratio is valid for any binning scheme, so either should work.  Smooth binning is statistically advantageous in that they make no \textit{a priori} assumptions regarding the data and generally produce more degrees of freedom in the fit (and is used in this work); irregular binning can better account for known errors or uncertainties in the data or theoretical isochrones.

\subsection{Incorporation of Outside Information \label{sec_outside}}

The fitting procedure described in this paper makes no \textit{a priori} assumptions regarding the distance, extinction, SFR(t,Z), etc.  While this allows a pure, unbiased estimate of these parameters, there are frequently constraints that should be applied in order to measure the star formation history as accurately as is possible.  The simplistic approach would be to limit the search space to correspond to the maximum allowable values of distance, extinction, metallicity, etc.  However, the incorporation of outside information can be done in a way more consistent with the ``maximum likelihood'' approach that is favored here.  Recalling that the fit parameter is merely $-2 \ln ( \hbox{probability} )$, one can factor in the probabilities of the trial star formation history matching other observational data in the same way.

For example, if the mean metallicity of a galaxy is known to be $\mean{\feh} = -1.5 \pm 0.3$ with Gaussian errors, but the trial fit has a mean metallicity of $\feh = -2.1$, one can add 4.0 to the fit parameter, thus giving the metallicities equal weight as the photometry.  (The value of four comes from the Gaussian likelihood ratio, $\chi^2$.)  In practice, there are several complications to using spectroscopic metallicity information.  First, the stars used in spectroscopic surveys (red giants, for example) are generally not representative of all stars in the galaxy.  Specifically, red giants are older than upper main sequence stars, so using $\mean{\feh}$ of ages older than 2 Gyr may be more appropriate than a $\mean{\feh}$ of all ages.  Second, most objects show significant spreads in metallicity; it would be preferable to compare the histograms of metallicities rather than just the means.  Finally, one must be careful about what one means by ``metallicity''.  The metallicity used in the models is [M/H]; that coming from spectroscopic studies is generally the abundance of one or more elements, such as [Fe/H] or [Ca/H], and is not necessarily the same value.

However, after accounting for all of these possibilities, constraints from spectroscopic studies, variable star distance measurements, extinction maps, etc. can (and should) be added to the fit parameter, producing a combined fit parameter.  This will allow one to answer not only the question of ``how well does the star formation history match the present photometry'', but ``how well does the star formation history match all known information about this galaxy.''  Answering second question clearly provides more stringent constraints on the solution than the first.

\section{APPLICATION TO DATA}

Having detailed a method for the determination of star formation histories, we now study its application to data sets, both simulated and real.  The tests with simulated data will examine how well star formation histories can be measured when the isochrones are a perfect match to the data; the tests with real data allow us to examine the effects of possible errors in the theoretical isochrones.  Given that there are certainly errors at some level, the question that must be answered is whether or not one can obtain a reasonable star formation history given these very good but imperfect theoretical models.  The Galactic dwarf spheroidal companions provide ideal targets for this test, because they (1) are sufficiently close that the ancient main sequence turnoff (MSTO) is visible, (2) are sufficiently far that line-of-sight depth is not a large problem, (3) have little dust to cause internal reddening, and (4) the majority have relatively simple star formation histories.

The procedure used to measure star formation histories was identical to that described in section 2, except that no incorporation of outside data (as described in section \ref{sec_outside}) was made.  This limitation was intentional, as the purpose of these tests is to determine the capabilities of CMD analysis alone.  (At any rate, given that the WFPC2 field of view is much smaller than the galaxies, we cannot expect to obtain star formation histories as good as those obtained from wide-field ground-based images.)

\subsection{Synthetic Galaxy 1: Single-Population \label{sec_sim1}}

\begin{figure}
\epsscale{0.95}
\plotone{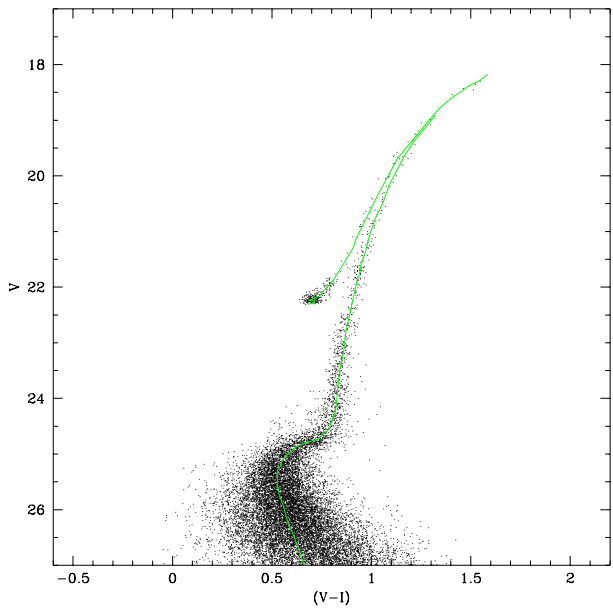}
\caption{CMD of Synthetic Galaxy 1.  The isochrone corresponds to the mean age and metallicity of the galaxy, 11.5 Gyr and $\logz = -1.7$.}
\label{fig_cmdsyn1}
\end{figure}

The first test of the method is an attempt to reconstruct a simple single-burst population, with the CMD shown in Figure \ref{fig_cmdsyn1}.  To provide a reasonable comparison to Leo II, a true distance modulus of $21.60$ with zero extinction, as well as the Leo II artificial star library, was used when generating the synthetic data.  The stars were distributed evenly in age between 11 and 12 Gyr, and in metallicity between $\logz = -1.75$ and $-1.65$ -- here and elsewhere, $\logz \equiv \log \frac{Z}{0.02}$ on the scale of the Girardi et al. (2000) isochrones -- effectively creating a single-population system (in terms of galaxy field populations).  The total number of stars in the CMD is 16449.  This galaxy will also serve as an illustration of the analysis procedure.

\begin{figure}
\plotone{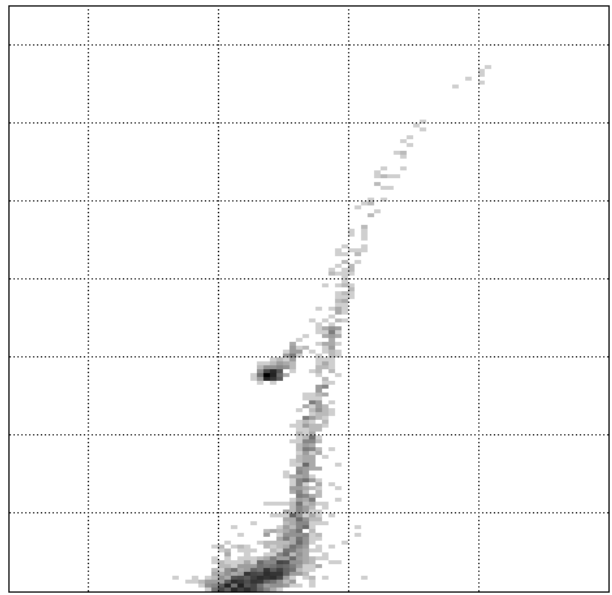}
\caption{``Observed'' CMD of synthetic galaxy 1, shown in greyscale.  The limits on the plot are those used in the solution: $17.5 < V < 25.0$ and $-0.3 < V-I < 2.0$.}
\label{fig_obssyn1}
\end{figure}

The first set of decisions that must be made is the CMD region to study and the binning size.  In order to retain information regarding the oldest MSTO stars while eliminating the stars with the worst photometric error, faint-end cuts of $V < 25.0$ and $I < 24.5$ will be used in this solution.  On the bright end, we cut off where the incompleteness due to saturation reaches 50\%, which gives requirements of $V > 17.5$ and $I > 16.5$.  A CMD binning size of 0.05 magnitudes in $V$ by 0.025 magnitudes in $(V-I)$ is sufficiently small to ensure that all CMD information is retained.  (The $1 \times 2$ shape of the rectangles was chosen to give similar sensitivity in both distance and extinction, as $E(V-I)/A_V$ is slightly more than 0.4.)  The ``observed'' CMD, binned accordingly, is shown in Figure \ref{fig_obssyn1}.

\begin{figure}
\plotone{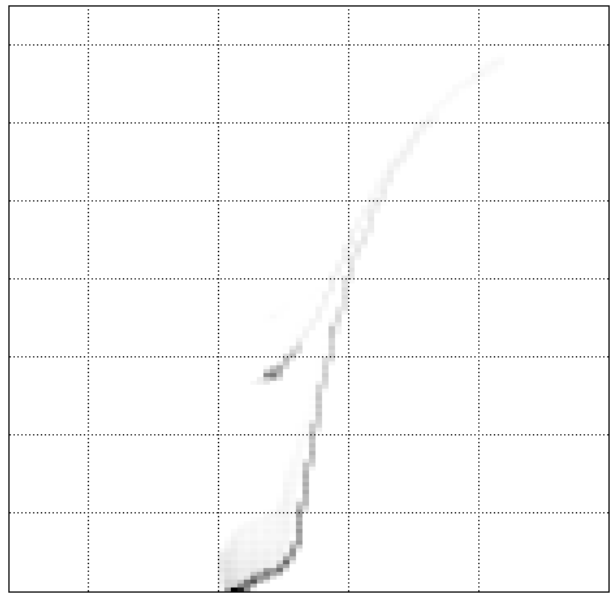}
\caption{Pure model CMD (shown in greyscale), calculated with an age of $11-12$ Gyr and metallicity of $\logz = -1.75$ to $-1.65$.  The limits on the plot are those used in the solution: $17.5 < V < 25.0$ and $-0.3 < V-I < 2.0$.}
\label{fig_puresyn1}
\end{figure}

\begin{figure}
\plotone{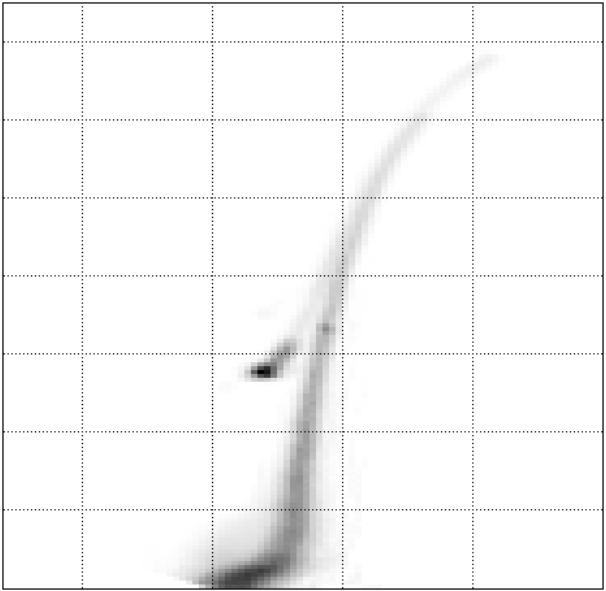}
\caption{Partial CMD (shown in greyscale), calculated with an age of $11-12$ Gyr and metallicity of $\logz = -1.75$ to $-1.65$.  The limits on the plot are those used in the solution: $17.5 < V < 25.0$ and $-0.3 < V-I < 2.0$.}
\label{fig_partsyn1}
\end{figure}

The second set of decisions will be the parameter space to explore.  Because we are attempting to ascertain the degree of accuracy with which a galaxy's star formation history can be recovered, we will attempt a solution with very high resolution (higher than will be used below when studying the real galaxies).  The metallicity stepsize will be $0.1$ dex in $\logz$, the age stepsize will range from 0.3 Gyr at young ages to 1 Gyr at old ages, and the distance modulus and extinction ($A_V$) will both be solved with a resolution of 0.02 magnitudes.  Because we have not retained any of the lower main sequence, it will be impossible to determine the IMF or binary distribution; these values have been fixed at a Salpeter slope and a binary fraction of 40\% with flat secondary mass function.  Using these parameters, there are 19 time bins and 19 metallicity bins, for a total of 361 partial CMDs.  The pure partial CMD corresponding to the input age and metallicity is shown in Figure \ref{fig_puresyn1}.  After application of observational errors (from the artificial star tests), the pure partial CMD becomes the final partial CMD shown in Figure \ref{fig_partsyn1}.

A comparison of Figures \ref{fig_obssyn1} and \ref{fig_partsyn1} shows that the simulated observed CMD was indeed drawn from the model CMD, in that the overall shape and density of points are the same, and no bin with a model value of zero has a nonzero number of observed points.  (The last point is not obvious from the printed images, given the limitations of greyscaling.)  Because of this, it is possible to make a statistically-valid fit of the simulated observations given the ensemble of partial model CMDs.  Solving for SFR(t,Z) at a variety of distance/extinction combinations, I was able to measure a minimized fit parameter and its corresponding distance, extinction, and star formation history.

\begin{figure}
\plotone{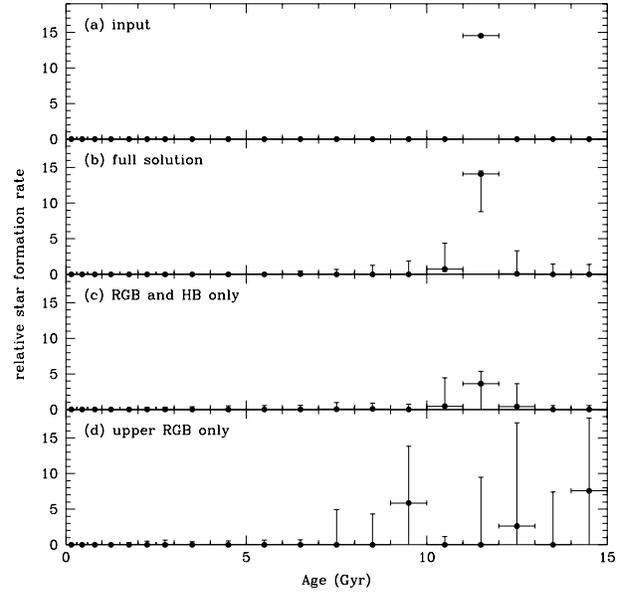}
\caption{Star formation histories of Synthetic Galaxy 1.  Panel a is the input history, panel b is the measured history using the entire CMD, panel c is the measured history without the turnoff, and panel d is the measured history with only the upper RGB.  Rates are given relative to the lifetime average rate of $3.44 \times 10{-5} M_\odot yr^-1$.}
\label{fig_sfhsyn1}
\end{figure}

The number of effective free parameters in the solution is 9 -- 7 star formation rates returned non-zero values, plus distance and extinction.  The mean difference between the fit parameters of the best fit and underlying star formation history is 9.8, meaning that the error bars are given by all star formation histories whose fit parameters are within 9.8 of the minimum (1088.1).  The distance ($\dist = 21.60 \pm 0.02$) and extinction ($A_V = 0.00 \pm 0.01$) can be immediately determined based on the best fits at each distance and extinction; both values agree with the input values.  The measured star formation rate is shown in panel b of Figure \ref{fig_sfhsyn1}, and matches the input star formation history extremely well.  At the $1\sigma$ level, there has been a small amount of bleeding from the $11-12$ Gyr bin into adjacent bins; however the input and recovered histories are consistent at the $2 \sigma$ level and the bleeding amounts to a loss of only 3\% of the star formation in the peak bin.  The metallicity was measured correctly, with a determined value of $\logz = -1.70 \pm 0.05$ dex.

In order to determine the quality of the fit, the expectation value of the fit parameter (1089.9) and its expected variance (1937.1) must be calculated.  Comparing with the corrected fit parameter value of 1097.9, this translates into a goodness-of-fit value of $Q = 0.18$, or $\chi^2_{eff} = 1.01$, meaning a statistically consistent fit to the ``observed'' data.  (In terms of percentiles, the fit is consistent at the 44\% level, meaning that it is better than 44\% of random drawings.)

The chance of having a good fit is enhanced, of course, as the binning of both age and metallicity matches that used when creating the model.  Whether or not a bad choice of bins affects the star formation history can be tested by running a solution using a different binning scheme.  By solving with the logarithmic age scheme used for the observed galaxies, one gets a worst-case estimate (as the break between the oldest two bins falls at the age of this system).  In making this test, I found a very poor fit quality ($Q = 7.19$ and $\chi^2_{eff} = 1.37$), but measured the correct distance, extinction, and star formation history.  The conclusion is thus that binning choices can hurt the fit quality, but are unlikely to affect the measured values or their uncertainties.  This result is not entirely a surprise, as the nature of the maximum likelihood ratio causes the solution to attempt to match all observed points with model points, even if this causes other model points to fall where observed points do not.  (For example, finding one star where the model predicts zero stars has a probability of 0.0, while finding no stars where one is predicted has a probability of 0.37.)  Thus all component populations will be fit.

In order to apply this test to more distant galaxies, it is also important to answer the question of how well the input star formation history can be recovered in more distant galaxies where the ancient main sequence turnoffs are not present.  To accomplish this, the star formation histories were calculated using photometric cutoffs brighter by 1.5 and 3.5 magnitudes in both $V$ and $I$.  Increasing the cutoffs by 1.5 magnitudes limits the solution to the RGB and HB; increasing by 3.5 limits it to just the upper RGB.  The measured star formation histories from these solutions are shown in panels c and d of Figure \ref{fig_sfhsyn1}.

The solution with the full RGB and horizontal branch (HB) was extremely successful, measuring the distance ($\dist = 21.60 \pm 0.02$), extinction ($A_V = 0.00 \pm 0.01$), star formation history, and mean metallicity ($\logz = -1.70 \pm 0.05$) with nearly the same accuracy as the fit to the entire CMD.  In fact, increasing the number of stars by a factor of 4 to compensate for the 1.5 magnitudes lost (1.5 magnitudes of distance modulus equals a factor of 2 in distance) would have produced a final solution to the same level of accuracy as the full fit.  The reason for the equally-accurate solution is threefold.  First, despite losing the main sequence turnoff, nearly all evolved stars remain above the photometric cutoff.  Second, age and metallicity are not completely degenerate on the RGB, allowing a sensitive numerical fit to ascertain the correct star formation history.  Finally, the HB morphology is sensitive to age, which also helps to break the degeneracy.  The last reason is as much a help as a hindrance, however, as the theoretical isochrones generally have much greater systematic uncertainties in the HB than in the RGB.

The solution measuring the upper RGB alone lost a great deal of information.  Although the distance and extinction measurements were both accurate ($\dist = 21.64 \pm 0.15$ and $A_V = 0.00 \pm 0.03$) they were not precise.  Increasing the number of stars by a factor of 25 to compensate for the greater distance would reduce the distance and extinction uncertainties, but would not greatly improve the star formation history measurement.  Specifically, with only the upper RGB, an acceptable fit can be obtained with stars of any age between 7 and 15 Gyr, due to the near-degeneracy of age and metallicity.

\subsection{Synthetic Galaxy 2: Composite-Population \label{sec_sim2}}

\begin{figure}
\plotone{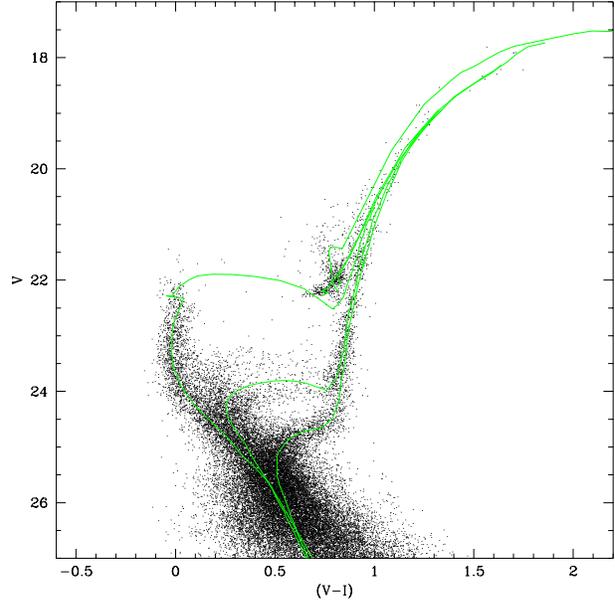}
\caption{CMD of Synthetic Galaxy 2.  The isochrones correspond to the mean ages and metallicities of the three bursts.  The youngest is 0.8 Gyr and $\logz = -1.0$, the middle is 3.6 Gyr and $\logz = -1.3$, and the oldest is 10.0 Gyr and $\logz = -1.6$.}
\label{fig_cmdsyn2}
\end{figure}

The second synthetic galaxy, whose CMD is shown in Figure \ref{fig_cmdsyn2}, is a system with a more complex star formation history.  The same input parameters were used, except for the metallicity and age distribution.  As is clear from the CMD, the star formation history consists of three bursts:
\begin{enumerate}
\item 0.6 to 1.0 Gyr, $\logz = -1.0$, $\sim 5000$ stars
\item 2 to 5 Gyr, $\logz$ increasing from $-1.4$ to $-1.2$, $\sim 11000$ stars
\item 8 to 13 Gyr, $\logz$ increasing from $-1.7$ to $-1.6$, $\sim 16000$ stars
\end{enumerate}
This star formation history was chosen to maximize the possibility of error in the solution.  The metallicity enrichment law used here exactly matches the age-metallicity pseudo-degeneracy, in that the RGB colour is the same at for stars of all ages.  (This is the reason for the metallicity skips between the bursts.)  Likewise, the system does not contain stars of extreme ages (0 or 15 Gyr), allowing the solution to err in finding such stars.  Finally, the three bursts have different shapes.  The young burst has a constant star formation rate, the middle burst has the lowest rate in the middle of the burst, and the oldest burst has highest rate in the middle.  The total number of stars in the CMD is 32518.
\begin{figure}
\plotone{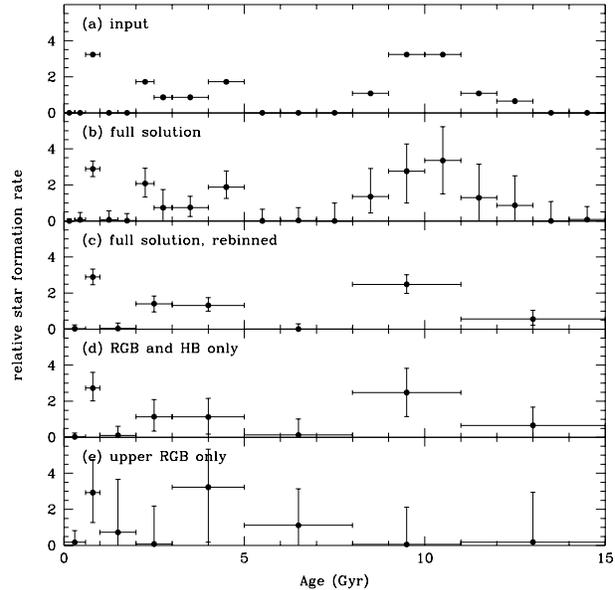}
\caption{Star formation histories of Synthetic Galaxy 2.  Panel a is the input history, panel b is the measured history using the entire CMD, panel c is the measured history using the entire CMD, rebinned for smaller uncertainties, panel d is the measured history without the turnoff, and panel e is the measured history with only the upper RGB.  Rates are given relative to the lifetime average rate of $4.68 \times 10{-5} M_\odot yr^-1$.}
\label{fig_sfhsyn2}
\end{figure}

The recovered star formation history, given in panels b and c of Figure \ref{fig_sfhsyn2}, has a minimized fit parameter of 2990.3.  Because of the more complex star formation history of this galaxy, the number of free parameters in the fit was larger (37 star formation rates plus distance and extinction = 39), which produced a maximum acceptable fit parameter of 3032.7.  The fit quality is not excellent ($Q = 1.80$ and $\chi^2_{eff} = 1.06$), but is consistent at better than a $2 \sigma$ (5\%) level.

\begin{table*}
\begin{center}
\caption{Observations}
\label{tab_observations}
\begin{tabular}{cccccc}
\hline
Galaxy          & Prog ID  & Date           & F555W Exposures                & F814W Exposures              & reference \\
\hline
Carina          & GTO 5637 & Jan 1995       & 200s, 2$\times$1100s           & 200s, 2$\times$1100s         & PI Westphal \\
Draco           & GTO 6234 & Jun 1995       & 200s, 2$\times$1000s$^a$       & 200s, 1100s, 1300s           & Grillmair et al. 1998 \\
Leo I           & GO 5350  & Mar 1994       & 350s, 3$\times$1900s           & 300s, 3$\times$1600s         & Gallart et al. 1999a \\
Leo II          & GO 5386  & May 1994       & 2$\times$80s, 8$\times$600s    & 2$\times$80s, 8$\times$600s  & Mighell \& Rich 1996 \\
Sagittarius$^b$ & GO 6614  & May$-$Oct 1996 & 6$\times$160s, 2$\times$600s   & 5$\times$160s, 2$\times$500s & Mighell et al. 1997 \\
Sculptor        & GTO 6866 & Dec 1997       & 2$\times$1200s, 2$\times$1300s & 4$\times$1300s               & Monkiewicz et al. 1999 \\
Ursa Minor      & GTO 6282 & Jul 1995       & 200s, 2$\times$1100s$^a$       & 200s, 2$\times$1100s         & PI Westphal \\
\hline
\end{tabular}
\end{center}
$^a$F606W was used for Draco instead of F555W \\
$^b$Six pointings were obtained in Sagittarius, comprising of three pairs of partially-overlapping fields.  One pair is $0.2^{\circ}$ from the centre, the second is $2.4^{\circ}$ from the centre, and the third is a field containing only Galactic foreground stars.
\end{table*}

From an examination of the history in panel c of Figure \ref{fig_sfhsyn2}, it is clear that there are rather large uncertainties in the star formation rates caused by the ability of of the prolonged star formation episodes to be modeled acceptably using different combinations of populations.  (This was not an issue for galaxy 1, since there was only a single burst.)  For example, the star formation measured in the $2.5-3$ Gyr bin, while a correct measurement of the input value, can be moved to the adjacent bins without significantly hurting the quality of the fit.  Thus while the input star formation history was recovered correctly, we do not have the ability to recover the structure of the bursts with a high signal-to-noise.  In order to compensate for this lack of time resolution in the bursts, panel c of Figure \ref{fig_sfhsyn2} shows the star formation history after additional binning in a somewhat logarithmic scheme.  As mentioned in section \ref{sec_unc}, the error bars tend to drop by roughly a factor of 2 (rather than by $\sqrt{2}$) when combining two bins, because errors in adjacent bins are correlated.  As with Synthetic Galaxy 1, the metallicity was measured very well, to an accuracy of rougly 0.15 dex in highly-populated age bins and 0.3 dex in less-populated bins.

As with galaxy 1, I have run additional star formation history solutions with photometric cutoffs that are brighter by 1.5 and 3.5 magnitudes.  These results are shown in panels d and e of Figure \ref{fig_sfhsyn2}.  Once again, the ability to measure the distance ($\dist = 21.59 \pm 0.05$), extinction ($A_V = 0.00 \pm 0.03$) and star formation history is essentially undiminished when subtracting 1.5 magnitudes from the photometric cutoff.  Even without increasing the number of stars (the 1.5 magnitudes of loss again corresponding to a factor of 2 in distance, or a factor of 4 in number of stars in the field of view), all input parameters were correctly measured.  However, the solution with 3.5 magnitude lost was poorly constrained, with large amounts of star formation again falling into adjacent bins.  All values were measured accurately given the uncertainties, of course, but the uncertainties were extremely large.

The conclusion from the limited-photometry solutions is that, while it is always preferable to have photometry reaching to ancient main sequence turnoffs, star formation histories can be measured accurately when the photometry reaches only $M_V = +2$.  With more restricted photometry, however, only broad features of the star formation history (perhaps a time resolution of $\log t = 0.5$) can be obtained.

\subsection{Real Galaxy: Leo II}

After analysing the pair of synthetic galaxies, we finally turn our attention to the measurement of star formation histories of a real galaxy.  Leo II will provide the primary example, because it is the only system in this sample with many stars and a primarily-old star formation history.

The data were all obtained from the Hubble Space Telescope archive using OTFC; all data sets are non-proprietary.  The list of observations is given in Table \ref{tab_observations}.  The data were cosmic-ray cleaned and combined using the \textit{crclean} algorithm of HSTphot (Dolphin 2000b), which is able to combine exposures at different gain settings, to provide one deep F555W image and one deep F814W image.

HSTphot was then used to obtain stellar photometry and artificial star tests using 43600 artficial stars.  Stars (both real and artificial) were required to have $\chi < 2.5$, $\hbox{S/N}> 5$, and $|\hbox{sharpness}| < 0.3$ in order to be considered detections.  Aperture corrections were determined separately for each chip and image, and are accurate to half a percent (the typical solution had 45 bright stars with an rms scatter of 0.03 magnitudes).  CTE loss corrections and transformations to $VI$ were made using an updated calibration solution from that given by Dolphin (2000c); the new equations are available on the author's web site.

\begin{figure}
\plotone{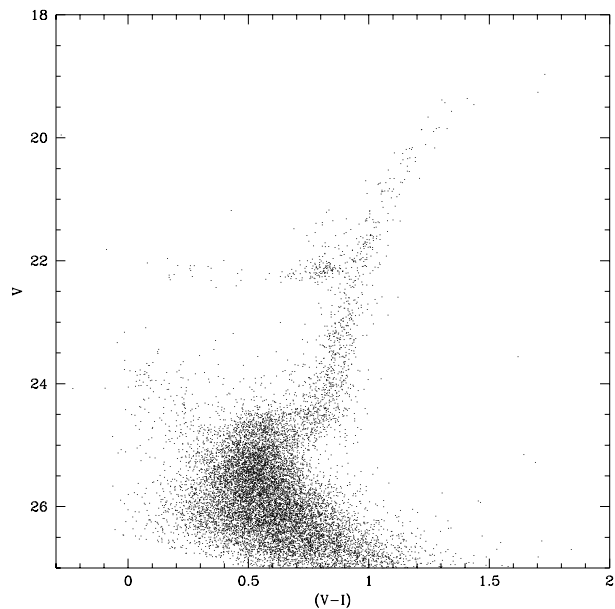}
\caption{Observed $(V-I)$,$V$ CMD of Leo II; N=12642; N($M_V<4$)=5188.  The overplotted isochrone corresponds to the mean values of the best solution: $\feh = -1.13$ and $t = 9.44$ Gyr.}
\label{fig_leoIIcmd}
\end{figure}

The Leo II CMD is shown in Figure \ref{fig_leoIIcmd}.  The 50\% completeness regimes, determined using artificial star tests, are $15.9 < V < 27.1$ and $14.8 < I < 26.1$ (the bright limit determined by saturation and the faint limit by loss of photons).

\begin{table*}
\begin{center}
\caption{Distance and Extinction Values}
\label{tab_distance}
\begin{tabular}{cccccccc}
\hline
                      & \multicolumn{3}{c}{Literature} & \multicolumn{2}{c}{Semi-Empirical} & \multicolumn{2}{c}{CMD Fitting} \\
Galaxy                & $\dist^a$  & $A_V^b$  & $\logz^c$   & $\dist$  & $A_V$   & $\dist$  & $A_V$ \\
\hline
Carina                & $20.03\pm0.09$   & 0.20 & $-2.0\pm0.2$   & $19.94\pm0.20^d$ & $0.19\pm0.10^e$ & $20.19\pm0.13$ & $0.00\pm0.14$ \\
Draco                 & $19.58\pm0.15$   & 0.08 & $-2.0\pm0.15$  & $19.60\pm0.18^f$ & $0.27\pm0.15^e$ & $19.49\pm0.11$ & $0.28\pm0.08$ \\
Leo I                 & $21.99\pm0.20$   & 0.11 & $-1.5\pm0.4$   & $21.84\pm0.14^d$ & $0.02\pm0.05^e$ & $21.80\pm0.06$ & $0.04\pm0.05$ \\
Leo II                & $21.63\pm0.09$   & 0.06 & $-1.9\pm0.1$   & $21.67\pm0.10^f$ & $0.12\pm0.07^e$ & $21.55\pm0.08$ & $0.00\pm0.09$ \\
Sagittarius (central) & $17.20\pm0.15^g$ & 0.47 & $-0.3\pm0.2^h$ & ...              & ...             & $17.11\pm0.14$ & $0.46\pm0.11$ \\
Sagittarius (outer)   & $17.20\pm0.15^g$ & 0.37 & $-0.3\pm0.2^h$ & ...              & ...             & $17.09\pm0.17$ & $0.45\pm0.13$ \\
Sculptor              & $19.54\pm0.08$   & 0.06 & $-1.8\pm0.1$   & ...              & ...             & $19.45\pm0.31$ & $0.06\pm0.19$ \\
Ursa Minor            & $19.14\pm0.10^i$ & 0.10 & $-2.2\pm0.1$   & $19.28\pm0.25^f$ & ...             & $19.16\pm0.11$ & $0.12\pm0.09$ \\
\hline
\end{tabular}
\end{center}
$^a$Except where noted, distances were taken from Mateo's (1998) compilation of literature values.\\
$^b$Calculated using the maps of Schlegel, Finkbeiner, \& Davis (1998) \\
$^c$Except where noted, metallicities were taken primarily from Mateo's (1998) compilation of literature values of $\feh$.\\
$^d$Measured using both the RGB tip, calibrated with the Girardi et al. (2000) isochrones, and the red clump technique described by Dolphin et al. (2001b) and Girardi \& Salaris (2001). \\
$^e$Measured using the RGB color, as per Sarajedini (1994). \\
$^f$Measured using the horizontal branch magnitude, with the calibration of Carretta et al. (2000). \\
$^g$From Bellazzini, Ferraro, \& Buonanno (1999a) \\
$^h$From Alard (2001), Bonifacio et al. (2000), and Cole (2001). \\
$^i$Includes a more recent measurement by Mighell \& Burke (1999). \\
\end{table*}

Table \ref{tab_distance} lists literature values of the distance, extinction, and RGB metallicity, as well as semi-empirical measurements of the distance and extinction calculated from my CMDs (when possible).  A sanity check on the photometric calibration is that the values agree; in the case of Leo II this is true.  Table \ref{tab_distance} also lists the values obtained from the CMD-fitting algorithm; a sanity check on the fit is the agreement between the semi-empirical and CMD-fitting values.

The CMD of Leo II contains a total of 12642 stars, of which 5188 are brighter than $M_V = +4$ ($V = 25.6$) and are thus useful in measuring the star formation history.  With the brightest stars in an ancient, metal-poor population falling near $M_V = -3$ ($V = 18.6$), the photometry limits thus encompass the necessary range.  The CMD shows the basic features of an old population -- a strong horizontal branch and weak upper main sequence.  However, the width of the main sequence turnoff region, presence of main sequence stars above the turnoff, and stars in the red clump region (above the red horizontal branch) all indicate the presence of a younger stellar population as well.

The star formation history analysis technique used for the two synthetic galaxies was then applied to Leo II.  The minimized fit parameter was 2335.9, with acceptable values as up to 2374.9.  The fit quality parameters were $Q = 2.15$ and $\chi^2_{eff} = 1.09$, meaning that the fit is acceptable at the $2.15 \sigma$ level.

While the synthetic galaxies were constructed purely for the purpose of testing the method, we hope to obtain some basic scientific results from the studies of the real galaxies.  The age resolution for the Leo II solution was logarithmic, since isochrones are more evenly spaced in $\log t$ than in $t$.  In order to reduce the number of free parameters, these solutions used a metallicity resolution of 0.15 dex instead of 0.1 dex and 11 time bins rather than 19.  However, because of the possibility of foreground stars and bad stars, we included all three sources of contamination -- a foreground star CMD, a bad star CMD consisting of a completely random distribution, and a bad star CMD consisting of the observed data smoothed with a Gaussian kernel with $\sigma = 0.2$ magnitudes in $(V-I)$ and 0.4 magnitudes in $V$.  As should be clear from the lack of distant outliers in the CMD in Figure \ref{fig_leoIIcmd}, the first and second sources are negligible (and were measured to be zero), while a small contribution from the smoothed observed CMD was needed to produce the best fit.
\begin{figure}
\plotone{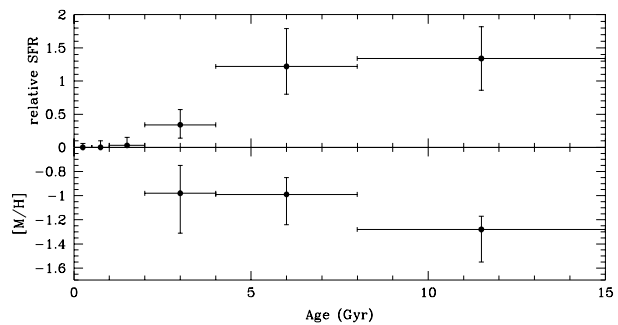}
\caption{Star formation and chemical enrichment histories of Leo II.  The top panel shows the star formation rate, normalized to the liftime average rate of $3.8 \times 10{-5} M_\odot yr^-1$.  The bottom panel shows the chemical enrichment history.  Although the solution was made with an age resolution of 0.15 dex, this and the following figures are plotted with a resolution of 0.3 dex to make the features clearer.}
\label{fig_sfhleoII}
\end{figure}

The measured star formation history is shown in Figure \ref{fig_sfhleoII}.  The obvious feature of this star formation history is a prolonged star formation epoch, lasting from 15 Gyr ago until about 5-6 Gyr ago.  The break at 5-6 Gyr is quite clear; the mean star formation rate at older ages is 7.3 times that at younger ages.  This finding is consistent with the star formation history recovered by Mighell \& Rich (1996), but is more extended (and older) than that found by Hernandez et al. (2000).  The reason for the discrepancy is unclear, but it should be noted that the younger history measured by Hernandez et al. (2000) cannot create the observed blue HB; thus some amount of older stars is necessary.  It is also apparent from the $1.75 \sigma$ detection of star formation in the $2-4$ Gyr bin that star formation extended until between 2 and 4 Gyr ago.

The measured metallicity values are surprisingly high.  The photometric metallicity measurement by Mighell \& Rich (1996), using the technique proposed by Sarajedini (1994) applied to these same data, gave a value of $\feh = -1.60 \pm 0.25$, while the preferred mean metallicity measured here is $-1.13_{-0.31}^{+0.09}$ -- a difference of nearly half a dex.  It should be noted, however, that the $\feh = -1.58$ red giant branch (M 2) of Da Costa \& Armandroff (1990) used by Mighell \& Rich has the same colour as an interpolated Girardi et al. (2000) isochrone of metallicity $\feh = -1.27$ and $\log t = 10.15$, consistent with the metallicities measured by Carretta \& Gratton (1997) for the similar-metallicity clusters NGC 3201, M 10, and NGC 6752.  Additionally, one must account for the fact that the mean age of a star in Leo II (9.4 Gyr) is much younger than that for a typical globular cluster; the same-colour isochrone at this age has a metallicity of $\feh = -1.21$.  Thus the metallicity measurements are entirely consistent; it is the scale used for calibration that is different.  That the present fit is internally-consistent is demonstrated by Figure \ref{fig_leoIIcmd}, which shows the interpolated isochrone corresponding to the measured distance, extinction, mean age, and mean metallicity.

In summary, the primary result is that the CMD-fitting algorithm produced the expected distance, extinction, and star formation history to within the uncertainties.  In specific, the extended star formation observed in other studies was recovered accurately, with $> 1 \sigma$ detections at old ages.  Additionally, a small amount of younger star formation was detected at the $>1 \sigma$ level.  Although the measured metallicity is much higher than values found in the literature, it is consistent once systematic differences in the calibrations have been corrected.

\section{SIX DWARF SPHEROIDALS}

In this section, I cover briefly the solutions for seven additional observed CMDs of six galaxies.  Each CMD will present a somewhat different challenge, with varying numbers of stars, complexity of star formation, and amount of foreground contamination.  All were reduced identically with HSTphot and the CMD analysis program.

\subsection{Draco}

\begin{figure}
\plotone{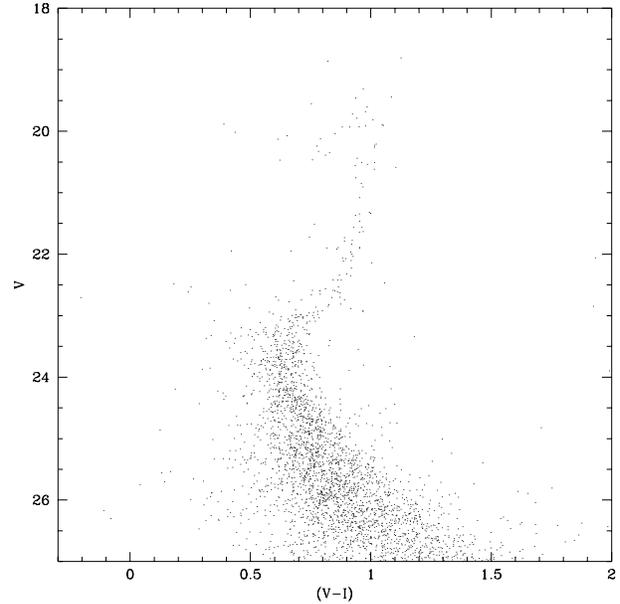}
\caption{Observed $(V-I)$,$V$ CMD of Draco; N=3371; N$(M_V<4)$=285.}
\label{fig_dracocmd}
\end{figure}

The first dwarf examined will be the Draco dwarf spheroidal.  From a cursory examination of the CMD (Figure \ref{fig_dracocmd}), one expects a simple (mostly-old) star formation history, as the turnoff and RGB are both extremely narrow.  A possible sign of trouble is that the semi-empirical extinction measurement is significantly greater than the literature value; this is possibly because of the filters used (F606W instead of the standard F555W).  However, a zero point error will affect the measured distance and extinction, but not the determined star formation history.  Note that stars within half a magnitude of the RGB tip would be saturated, but this should not significantly affect the CMD solution as the upper RGB is not strongly populated.

\begin{figure}
\plotone{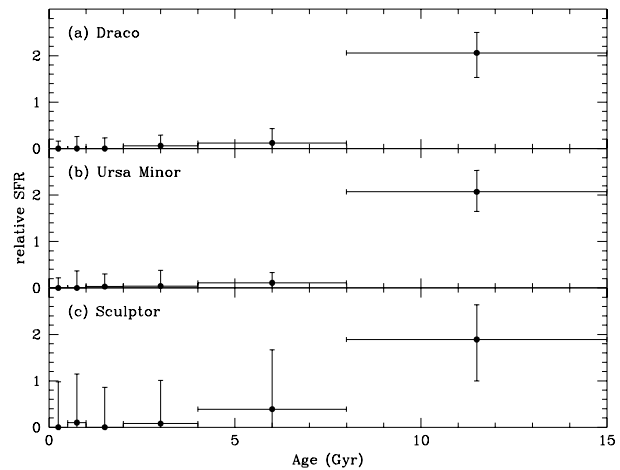}
\caption{Star formation histories of three old systems: Draco, Ursa Minor, and Sculptor.  Each is normalized relative to its lifetime average star formation rate.}
\label{fig_sfhdums}
\end{figure}

The measured star formation history is shown in panel a of Figure \ref{fig_sfhdums}.  As was expected from the visual examination of the CMD, the only significant star formation episode appears to be at ancient ages ($> 11$ Gyr ago).  The constraints on the maximum amount of younger star formation are given by the upper error bars; there is the possibility of a significant amount of star formation (more than the lifetime average rate) lasting until $8$ Gyr ago, but very little since then.  The mean metallicity is measured to be $\logz = -1.7 \pm 0.4$ dex.  The conclusion of an entirely ancient galaxy is consistent with that obtained by the other study of this data set (Grillmair et al. 1998) as well as the ground-based work of Carney \& Seitzer (1986).

\subsection{Ursa Minor}

\begin{figure}
\plotone{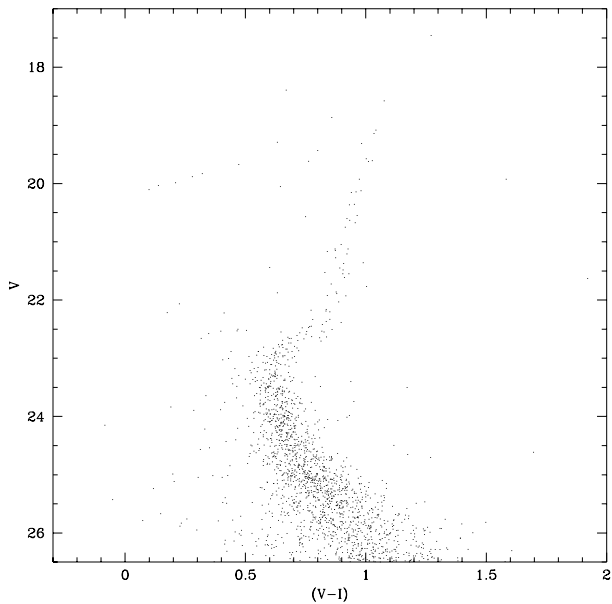}
\caption{Observed $(V-I)$,$V$ CMD of Ursa Minor; N=1941; N$(M_V<4)$=172.}
\label{fig_ursacmd}
\end{figure}

The Ursa Minor dwarf spheroidal is another system whose CMD contains a moderate number of stars.  Its CMD is shown in Figure \ref{fig_ursacmd}; as with Draco, a narrow turnoff and RGB imply a simple and predominantly-old star formation history.  A few stars exist above the turnoff, implying either the presence of blue stragglers or a small young population.  Note that stars within 0.2 magnitudes of the RGB tip would be saturated, but it is unlikely that many such stars exist in this small field.

The measured star formation history is shown in panel b of Figure \ref{fig_sfhdums}.  As was expected from the visual examination of the CMD, the only significant star formation episode appears to be at ancient ages ($> 11$ Gyr ago).  The mean metallicity was measured to be $\logz = -1.5 \pm 0.3$ dex.  The conclusion of an entirely ancient galaxy is consistent with that obtained by other studies of this data set (Mighell \& Burke 1999, Hernandez et al. 2000) -- although care should be taken in making too much of this comparison, as Hernandez et al. (2000) had an error of more than $-0.1$ magnitudes in their $(V-I)$ colours -- as well as the ground-based work of Olszewski \& Aaronson (1985).

\subsection{Sculptor}

\begin{figure}
\plotone{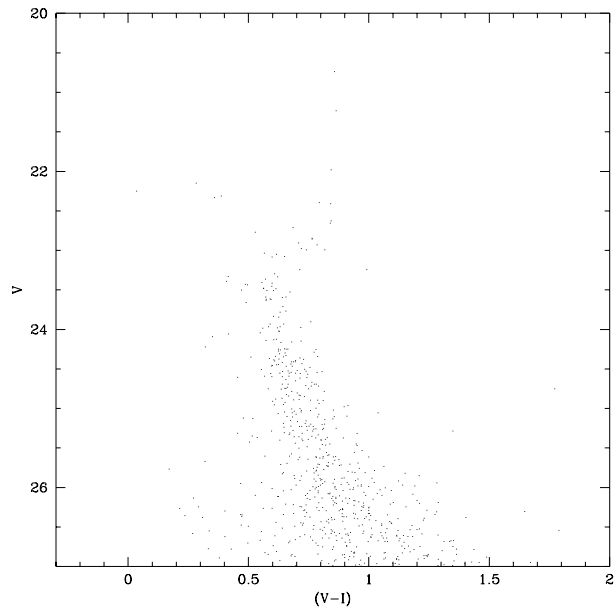}
\caption{Observed $(V-I)$,$V$ CMD of Sculptor; N=819; N$(M_V<4)$=46.}
\label{fig_sculptorcmd}
\end{figure}

The extreme case, in terms of numbers of stars, is that of the Sculptor dwarf spheroidal.  While the Leo II CMD had 5188 stars brighter than $M_V = +4$, Sculptor's (Figure \ref{fig_sculptorcmd}) has only 46.   The CMD cuts off at $M_V = +0.2$ because of saturation; thus the entire upper RGB is lost in these data.  Given the small number of evolved stars in the CMD, it is clearly impossible to measure the star formation history with great precision.  However, there the CMD shows no evidence of young stars.

The measured star formation history is shown in panel c of Figure \ref{fig_sfhdums}.  Consistent with the visual examination of the CMD, we again find an ancient galaxy.  The constraints on the maximum amount of younger star formation are quite weak, though, and it is possible to have an acceptable fit with any one of the younger low-resolution bins increased to the lifetime average star formation rate.  The mean metallicity is measured to be $\logz = -1.5 \pm 0.6$.  The conclusion of an entirely ancient galaxy is consistent with that obtained by other study of this data set (Monkiewicz et al. 1999) as well as the ground-based work of Da Costa (1984).

\subsection{Leo I}

\begin{figure}
\plotone{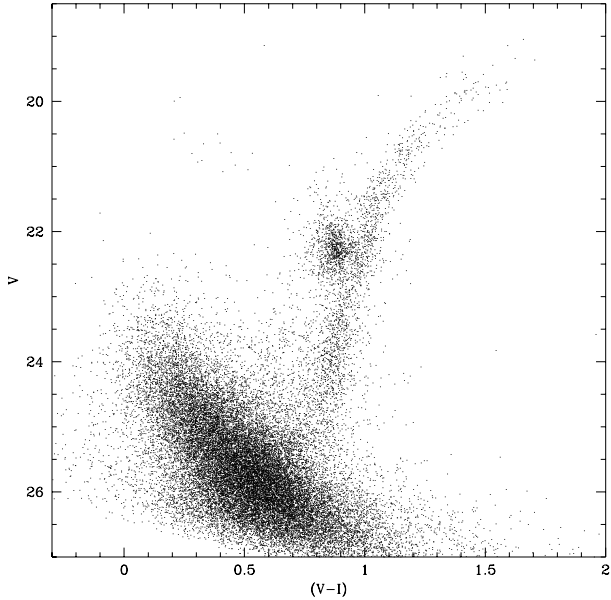}
\caption{Observed $(V-I)$,$V$ CMD of Leo I; N=31064; N$(M_V<4)$=22290.}
\label{fig_leoIcmd}
\end{figure}

Leo I is the first galaxy in this study to contain young stars.  Its CMD, shown in Figure \ref{fig_leoIcmd}, shows an extremely broad turnoff, ranging from ancient stars to very young stars.  The main sequence itself is dominated by a young population, and a few blue helium burners are present.  There is likely an HB extending bluewards from the base of the red clump; however one cannot be entirely sure that those are HB stars rather than young stars evolving off the main sequence.  It was the only object for which observations were made before WFPC2 was cooled; the CTE corrections and calibrations are thus somewhat more uncertain than for the other objects in this sample.  Nevertheless, as noted previously, any error in the zero points (there is no reason to believe any such error exists) would affect merely the distance and extinction measurements; the recovered star formation history is unaffected.

The goodness-of-fit was by far the worst of the galaxies studied.  The $Q$ value is 7.47, which is similar to that measured in the poorly-binned solution of synthetic galaxy 1; this may indicate that an ``unlucky'' choice of time bins.  Nevertheless, we can use the fact that the difference between the fit parameters of the best fit and that of the true fit is independent of the quality of the fit -- in other words, we can use the formalism defined in section \ref{sec_unc} to determine the uncertainties.

\begin{figure}
\plotone{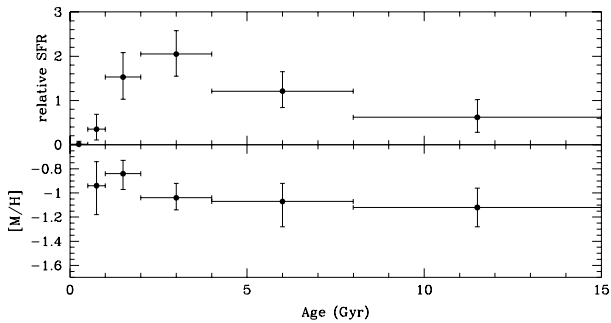}
\caption{Star formation and chemical enrichment histories of Leo I.  The top panel shows the star formation rate, normalized to the liftime average rate of $8.5 \times 10{-5} M_\odot yr^-1$.  The bottom panel shows the chemical enrichment history.}
\label{fig_sfhleoI}
\end{figure}

The measured star formation history is shown in Figure \ref{fig_sfhleoI}.  Leo I shows star formation detected at the $1 \sigma$ level at every age from 15 Gyr ago until 0.5 Gyr ago.  The largest epoch of star formation occurred recently, from 3 to 1 Gyr ago, during which time the star formation rate was 2.5 times the lifetime average.  The burst appears to have begun and ended quite quickly, as the star formation history is inconsistent with a constant rate from 4 Gyr ago until the present.  Additionally, ancient ($>11$ Gyr) stars are detected at the $1 \sigma$ level, despite confusion caused by the young population.

Nearly all previous studies using these data (Gallart et al. 1999b; Hernandez et al. 2000) have concluded that there was a strong episode of star formation recently in Leo I.  Gallart et al. (1999b) estimated that most of the star formation occurred between 1 Gyr and 7 Gyr ago; Hernandez et al. (2000) measured bursts centred near 4 and 7.5 Gyr.  However, the results of Hernandez et al. (2000) are skewed by an error of $\sim +0.2$ magnitudes in the photometric zero points, thus making the stars appear older and explaining why the two apparent peaks in my star formation history ($2-2.8$ and $4-5.7$ Gyr) are younger than those in theirs.

The study of the oldest stars from these data has had conflicting results.  Gallart et al. (1999b) found a negligible star formation rate beyond 12 Gyr, while Caputo et al. (1999) concluded that such star formation likely exists.  As noted above, I measure a presence of star formation older than 11 Gyr at the $1 \sigma$ level.  The Gallart et al. (1999b), result, however, is based on a very large assumed distance modulus of 22.18; bringing Leo I to the distance determined here would move some of their $9.4-12$ Gyr star formation into the $12-15$ Gyr range; the presence of old stars has been confirmed by NTT observations of the outer regions of Leo I by Held et al. (2000).

\subsection{Carina}

\begin{figure}
\plotone{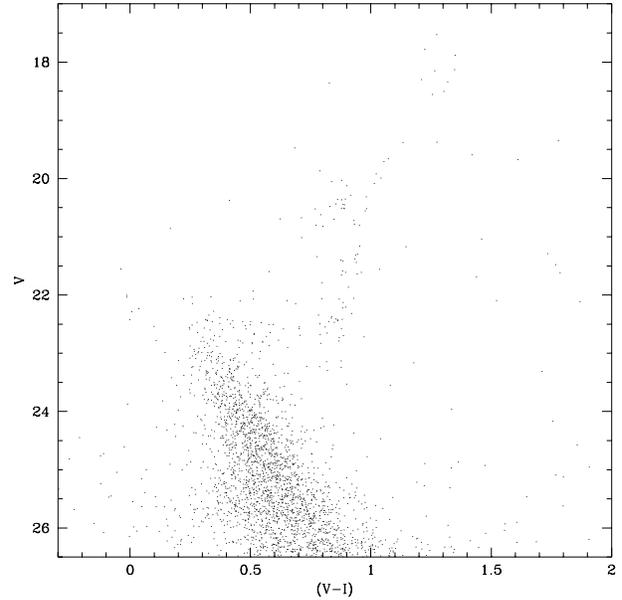}
\caption{Observed $(V-I)$,$V$ CMD of Carina; N=2772; N$(M_V<4)$=609.}
\label{fig_carinacmd}
\end{figure}

The Carina dwarf spheroidal is another system containing relatively young stars.  While its younger population does not dominate the CMD (Figure \ref{fig_carinacmd}) as much as that of Leo I, the significant foreground contamination makes the solution more difficult.  The main sequence extends to $V \sim 22.5$, corresponding to an absolute magnitude of $+2.3$.  The smaller number of stars makes stronger conclusions more difficult, of course -- the 15 blue helium burners seen in Leo I would scale down to 0.4 given the relative numbers of stars in the two fields.

The distance and extinction values from the CMD fit, shown in Table \ref{tab_distance}, are the only ones that disagree with semi-empirical values calculated from the CMD.  It is likely that the foreground contamination and lack of a well-photometered lower main sequence made the distance and extinction question poorly-constrained.  This is a system for which incorporation of outside information (such as the Schlegel et al. extinction value) would clearly be profitable.

\begin{figure}
\plotone{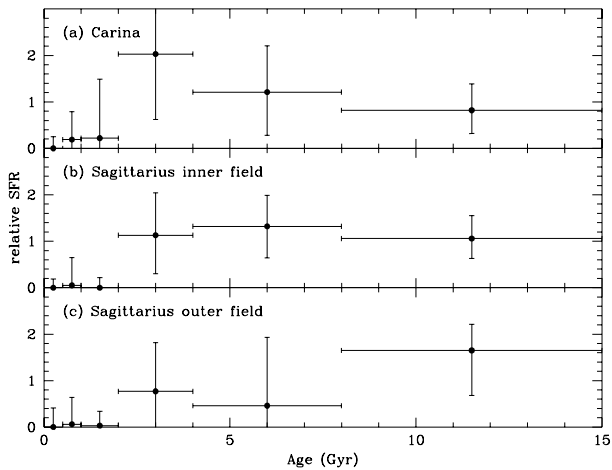}
\caption{Star formation histories of two mixed-age systems: Carina and Sagittarius.  Each is normalized relative to its lifetime average star formation rate.}
\label{fig_sfhcss}
\end{figure}

The measured star formation history is shown in panel a of Figure \ref{fig_sfhcss}.  Carina appears to show continuous star formation (at the resolution possible) from its earliest star formation episode until roughly 2 Gyr ago.  Optimal science can only be obtained with a wider-field camera, of course; the primary purpose of studying these data is to determine how precisely the star formation history can be determined from this CMD.

For comparison, the previous studies of these data (Mighell 1997, Hernandez et al. 2000) have also noted strong intermediate-aged stellar populations.  The bulk of stars found by Mighell (1997) have ages between 4 and 10 Gyr; Hernandez et al. (2000) measured 1 Gyr-wide peaks centred at 3, 5, and 8 Gyr.  Using ground-based data, Hurley-Keller, Mateo, \& Nemec (1998) find a 3-burst structure with ages 3, 7, and 15 Gyr.  The present work finds significant amounts of stars younger than 4 Gyr, contradicting Mighell (1997) but agreeing extremely well with Hurley-Keller et al. (1998).  As with Leo I, there is some ambiguity as to whether or not ancient stars exist.  Mighell (1997) and Hurley-Keller et al. (1998) found evidence of them, while Hernandez et al. (2000) did not.  Figure \ref{fig_sfhcss} shows the presence of old ($> 8$ Gyr) stars at the $1 \sigma$ level.  As was the case for Leo II, the presence of a strong blue HB (Smecker-Hane et al. 1994) would argue against the history proposed by Hernandez et al. (2000), who find essentially no star formation older than 10 Gyr.  A serious ($\sim -0.2$ magnitudes in $(V-I)$) error in their photometric zero point likely causes their spurious result.

\subsection{Sagittarius \label{sec_sag}}

\begin{figure}
\plotone{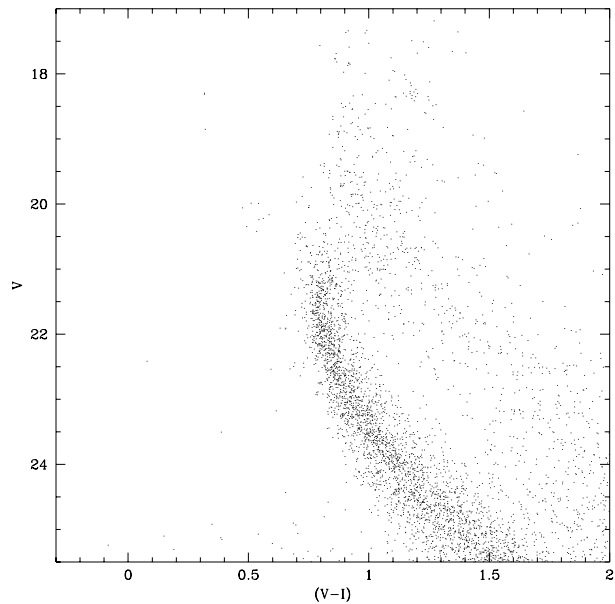}
\caption{Observed $(V-I)$,$V$ CMD of the central Sagittarius field; N=6553; N$(M_V<4)$=809.}
\label{fig_sag1cmd}
\end{figure}

\begin{figure}
\plotone{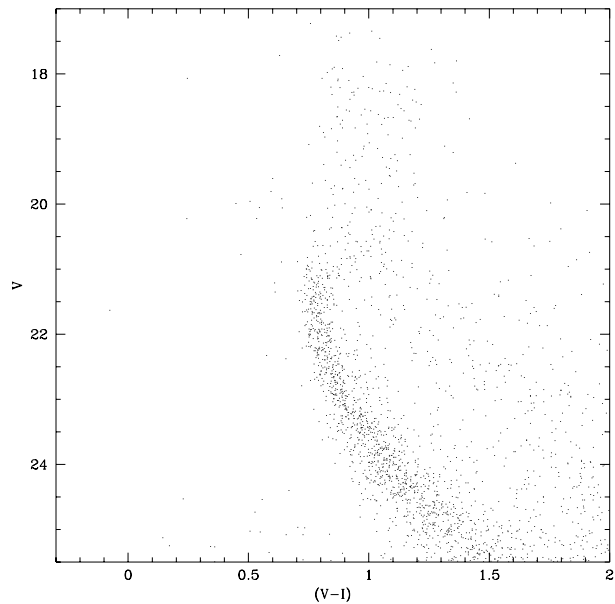}
\caption{Observed $(V-I)$,$V$ CMD of the outer Sagittarius field; N=3329; N$(M_V<4)$=408.}
\label{fig_sag2cmd}
\end{figure}

The final galaxy to be examined is the Sagittarius dwarf spheroidal.  It provides an even more difficult challenge than Carina, as the number of stars in the CMD is not significantly greater, while there is a tremendous amount of foreground contamination.  The CMDs of the central field ($0.2^{\circ}$ from the centre) and outer field ($2.4^{\circ}$ from the centre) are shown in Figures \ref{fig_sag1cmd} and \ref{fig_sag2cmd}.  Because of the foreground contamination, only a very rough estimate of the star formation can be made -- there are no extremely young stars, but the central field does appear to have a broad turnoff extending up to $V \sim 20.4$.  The turnoff in the outer field does not extend as high, but is still much broader than those of Ursa Minor or Draco.  In both fields, the CMD cuts off about 2 magnitudes below the RGB tip.  Given the foreground contamination, however, it is unlikely that Sagittarius RGB stars would have been distinguishable from foreground main sequence stars.

Because of the possibility of differences in extinction and star formation history, the CMDs of the two fields were fit separately.  The measured star formation histories are shown in panels b and c of Figure \ref{fig_sfhcss}.  The central field shows measurable star formation until $\sim 2$ Gyr ago, while the outer field shows only about half the number of young ($< 8$ Gyr) stars.  Sagittarius is also the only system for which significant chemical enrichment is measured.  The inner field shows a metallicity of $\logz = -1.1 \pm 0.4$ dex at ages older than 8 Gyr, which increased to $\logz = 0.0 \pm 0.2$ for the youngest large population of stars ($2-4$ Gyr old).  The history seen in the outer field is consitent, but the uncertainties are larger because of the smaller number of stars.

\begin{table*}
\begin{center}
\caption{Summary of Results}
\label{tab_summary}
\begin{tabular}{ccccccccc}
\hline
Quantity         & Ursa Minor     & Draco          & Sculptor       & Leo II         & Sagittarius    & Carina         & Leo I          \\
\hline
$\mean{t}$ (Gyr) & $12.7        $ & $12.2        $ & $11.5        $ & $ 9.4        $ & $ 9.0        $ & $ 7.2        $ & $ 6.1        $ \\
$\sigma t$ (Gyr) & $ 1.9        $ & $ 2.3        $ & $ 3.1        $ & $ 3.1        $ & $ 3.7        $ & $ 3.3        $ & $ 4.1        $ \\
$\mean{\logz}^a$   & $-1.5 \pm 0.3$ & $-1.8 \pm 0.4$ & $-1.5 \pm 0.6$ & $-1.1 \pm 0.3$ & $-0.6 \pm 0.4$ & $-1.2 \pm 0.4$ & $-1.0 \pm 0.2$ \\
\hline
\end{tabular}
\end{center}
$^a$Metallicities are given on the scale of the Girardi et al. (2000) isochrones; $\logz \equiv \log (Z/0.02)$.
\end{table*}

No literature data currently exists using these WFPC2 images; however a few ground-based studies have been carried out (Layden \& Sarajedini 1997; Marconi et al. 1998; and Bellazzini et al. 1999a).  Marconi et al. (1998) and Bellazzini et al. (1999b) found very large metallicity dispersions, consistent with the significant metallicity evolution measured in this work.  The star formation history is agreed to be extended, with a peak age agreed to fall between 8 and 11 Gyr; this compares favorably with the mean ages measured here (8.6 Gyr in the inner field and 9.8 Gyr in the outer field).

\section{CONCLUSIONS}

An examination of the technique for measuring star formation histories has been presented.  While the underlying concept -- finding the star formation history most likely to have produced the observed data -- is straightforward, there are a number of potential traps that must be overcome.  In generating synthetic CMDs, one must take care to ensure that all possible outcomes have been sampled; this is not done by ``random drawing'' techniques in which a certain number of stars are randomly drawn and placed on the CMD.  Instead, it is necessary to make a true model CMD -- a CMD that represents the probability distribution from which the data could have been drawn.  The first step is to make fine interpolations of the isochrones in age, metallicity, and mass so that all possible single stars are accounted for.  One must also account for the possibility of binaries by considering a number of possible secondary passes sufficiently large to create a smooth model CMD.  The final step in generating a partial CMD (CMD for a small range of age and metallicity) is to apply the results of artificial star tests.  The model CMD can be generated from any combination of the partial CMDs (different combinations correspond to different star formation histories), plus a model of foreground contamination and models of bad detections.

I have demonstrated the inadequacy of a $\chi^2$ minimization when fitting Poisson-distributed data (as is the case here).  Specifically, a $\chi^2$ minimization will always minimize with the wrong star formation history; the only question is how wrong the answer will be.  Instead, a Poisson likelihood ratio is recommended, the equation given in equation \ref{eq_plr}.  It has also been demonstrated that the ``Saha $W$'' (Saha 1998) statistic is not designed for model-data comparisons.  The Bayesian inference scheme of Tolstoy \& Saha (1996) provides an accurate solution of relative star formation rates but not the overall mean star formation rate.  The question of binning vs. non-binning is demonstrated to be unimportant, as the same star formation rate will be obtained so long as the bin sizes are as small as the smallest features of the model CMD.  Finally, techniques for measuring uncertainties and determining the overall fit quality are given, as well as a method in which outside data (such as a red giant metallicity distribution) can be incorporated into the fit without use of a prior.

The technique was then applied to a pair of synthetic galaxies -- one single-population and one composite-population.  The star formation history of the single-population system was measured with an age accuracy of $\pm 0.03$ dex and distance and extinction accuracy of 0.02 magnitudes, provided that at least the RGB and HB were included in the data (depth of $M_V = +2$).  Most of the constraints were lost, however, when restricting the solution to only the upper RGB (depth of $M_V = 0$); this introduced an age uncertainty of $\pm 0.2$ dex into the solution.  Although the quality of the fit was severely degraded when using an intentionally-wrong set of age bins, we note that the measured distance, extinction, and star formation history were all correct.  The star formation history of the synthetic composite-population system was measured with less accuracy, with resolution of roughly $\pm 0.07$ dex producing reasonable signal-to-noise with photometric depth of $M_V = +2$.  However, the solution with a photometric limit of $M_V = 0$ was again very uncertain, with age resolution degraded to $\pm 0.25$ dex.  The quoted resolutions, of course, are dependent upon the number of stars in the observed field; the uncertainties scale as $1/\sqrt{N}$.

Finally, I showed measurements of the star formation histories of seven dwarf spheroidal companions.  While each data set had a different quality (number of stars, photometric depth, and amount of foreground contamination), the ability to accurately measure uncertainties allows one to give the best answer and the uncertainty in the measurement for each object.  Thus the star formation history can always be measured -- even with the very poor Sculptor CMD -- but better data will naturally result in smaller uncertainties.

The technique-related findings of this study can be summarized as follows:
\begin{enumerate}
\item In every case, the calculated star formation history matched with the qualitative star formation history obtained by a cursory examination of the CMD.  In nearly every case, the distance and extinction were consistent with literature values.
\item The number of stars with $M_V < +4$ required to produce results with signal-to-noise $>$ 1 at moderate resolution appears to be about 150 for an old system and 500-1000 for a system with many young stars.
\item Even with the uncertainties in the isochrones, all CMDs were well-fit.  The largest $\chi^2_{eff}$ was 1.16, and only the Leo I fit was worse than 2.5 $\sigma$ from an ideal solution.
\item In the case of the Sagittarius dwarf, a large amount of foreground contamination (more foreground stars than Sagittarius stars) does not add significantly to the fit uncertainties.  This is likely because the main sequence and MSTO of Sagittarius are sufficiently separated from the bulge main sequence.
\end{enumerate}

Scientifically, the results are limited by the fact that only a small fraction of each galaxy was studied.  The Leo spheroidals had sufficient numbers of stars for accurate star formation history measurements; the others produced only rough star formation histories.  The consistent feature of the star formation histories is that ancient ($> 8$ Gyr) star formation was detected in all eight CMDs at the $1 \sigma$ level.  After the ancient burst, some (Ursa Minor, Draco, and Sculptor) show no evidence of young star formation.  Leo II shows star formation covering about half its lifetime, while Carina and Sagittarius appear to have formed stars until $\sim 2$ Gyr ago.  Finally, Leo I shows a very strong young burst, with its star formation rate $2-3$ Gyr ago nearly four times its lifetime average.  Results for the galaxies are summarized in Table \ref{tab_summary}.

\section*{Acknowledgments}
Support for this work was provided by NASA through grant number AR-09203.03 from the Space Telescope Science Institute.  All of the data presented in this paper were obtained from the Multimission Archive at the Space Telescope Science Institute (MAST). STScI is operated by the Association of Universities for Research in Astronomy, Inc., under NASA contract NAS5-26555.

\end{document}